\documentclass[twocolumn]{openjournal} 

\usepackage{comment}
\usepackage{lipsum}

\usepackage{mathtools}
\usepackage[breaklinks,colorlinks,citecolor=blue,urlcolor=blue,linkcolor=blue]{hyperref}

\newcommand{\high}{`high' }

\newcommand{\ps}{particle splitting }

\newcommand{\code}[1]{\texttt{#1}}
\newcommand{\epsff}{\epsilon_{\rm ff}}
\newcommand{\Msun}{\rm M_\odot}

\newcommand{\tff}{t_{\rm ff}}

\begin{document}

\title[Star Formation Mitosis]{Star Formation Mitosis: The Effects of Particle Splitting on the Star Formation Outcome}

\author[0000-0003-1053-1262]{Shivan Khullar}
\email{skhullar@flatironinstitute.org}
\affiliation{Center for Computational Astrophysics, Flatiron Institute, 162 Fifth Avenue, New York, NY 10010, USA}
\affiliation{David A. Dunlap Department of Astronomy \& Astrophysics, University of Toronto, 50 St. George St., Toronto, ON M5S 3H4, Canada}
\affiliation{Canadian Institute for Theoretical Astrophysics, University of Toronto, 60 St. George St., Toronto, ON M5S 3H8, Canada}

\author[0000-0002-1655-5604]{Michael Y. Grudi{\'c}}
\affiliation{Center for Computational Astrophysics, Flatiron Institute, 162 Fifth Avenue, New York, NY 10010, USA}

\author[0000-0001-9732-2281]{Christopher D. Matzner}
\affiliation{David A. Dunlap Department of Astronomy \& Astrophysics, University of Toronto, 50 St. George St., Toronto, ON M5S 3H4, Canada}

\author[0000-0003-3729-1684]{Philip F. Hopkins}
\affiliation{TAPIR, California Institute of Technology, Mailcode 350-17, Pasadena, CA 91125, USA}

\author[0000-0002-8659-3729]{Norman Murray}
\affiliation{Canadian Institute for Theoretical Astrophysics, University of Toronto, 60 St. George St., Toronto, ON M5S 3H8, Canada}

\author[0000-0002-4900-6628]{Claude-Andr{\'e} Faucher-Gigu{\`e}re}
\affiliation{CIERA \& Department of Physics and Astronomy, Northwestern University, 1800 Sherman Avenue, Evanston, IL 60201, USA}







\begin{abstract}

Particle splitting can increase the dynamic range in Lagrangian astrophysical simulations by several orders of magnitude. However, such methods can introduce numerical noise that may affect the rate of convergence, particularly for collapse problems where fragmentation could be artificially suppressed or enhanced. Thus it is necessary to determine how these errors propagate to key quantities of interest, and to correct or mitigate them where possible. Using \code{GIZMO}'s MFM-MHD method, we perform a set of $\approx 220$ simulations of a uniform-density turbulent gas sphere undergoing collapse, systematically varying the cloud mass and radius, feedback physics, initial resolution, the pace and cadence at which we split particles and the sink (accreting) or single-age stellar population (SSP; non-accreting) particle formation prescription. Crucially, we compare with control runs that start with uniformly-high resolution without splitting. While we find no evidence for artificial fragmentation caused by splitting particles, particle splitting runs have a smoother density field than their non-splitting counterparts. Due to this, particle splitting runs exhibit minor differences in the density and turbulence statistics. However, they form about the same total stellar mass as the control non-splitting runs, albeit with a slightly different mass spectrum. As a result, particle splitting runs have a different evolutionary history, compared to runs without particle splitting, especially when stellar feedback is present. Despite the changes seen in runs where we split particles, the numerical solution lies within the scatter inherently present in these chaotic systems, which we model by modifying the initial turbulent velocity field. Our results here argue that particle splitting generally ``works" for star formation applications, despite some systematic differences in the specifics.
\end{abstract}




\section{Introduction}

Simulating astrophysical systems has long posed a challenge due to the wide range of scales involved. Finite volume methods used on Eulerian grids have had success in many astrophysical applications \citep[e.g.][]{Klein1999a, Teyssier2002, Offner2009, Kim2013, Federrath2015b, Brucy2020, Agertz2021, Rosen2022, Polak2024, Beattie2025b}, in part due to Adaptive Mesh Refinement (AMR). Despite the use of AMR, the effective number of resolution elements ($N_{\rm eff}$) in these simulations rarely exceeds $N_{\rm eff} \sim (2^{13})^3 \sim 10^{11.5}$ or $\sim 4$ orders of magnitude in spatial scales \citep[e.g][]{Seifried2017, Colman2024}. This limitation is partly due to grid codes having to expend resolution to simulate relatively low density gas that may not be of interest. Re-simulating in multiple steps by taking initial conditions from a coarser resolution run can increase the overall dynamic range \citep[e.g.][]{Yang2025, Lewis2025, Mayer2025}. For example, \citet{Nordlund2014} achieve a $N_{\rm eff} \sim (2^{29})^3 \sim 10^{26}$ ($\sim 9$ orders of magnitude in spatial scales) while using RAMSES \citep{Teyssier2002} to simulate a protoplanetary disk in a $40 \, {\rm pc}$ GMC in multiple stages. 
Computational limits, however, imply that resolution cannot be increased indefinitely. In order to make the problem computationally tractable, the range of scales must be truncated at some level in both Eulerian and Lagrangian codes. In star formation calculations, this is typically done by introducing Lagrangian particles (sinks or star particles) that prevent the simulation timestep from reducing to computationally intractable values and thus grinding it to a halt \citep{Bate1995, Krumholz2004}.

Despite these practical computational limitations, Lagrangian codes that follow gas offer several advantages over Eulerian grid codes. Meshless finite mass or volume (MFM/MFV) methods in particular \citep{Gaburov2011}, conserve angular momentum, do not require artificial diffusion terms and do not suffer from the same truncation errors introduced while purely advecting gas in grid codes \citep{Hopkins2015}. They are also particularly advantageous for solving collapse problems since they do not introduce spurious artificial fragmentation \citep{Truelove1997, Grudic2021}. \cite{Hubber2006} show that 2 particle widths per Jeans length are sufficient to prevent artificial fragmentation, instead of the minimum of 4 cells per Jeans length used in AMR calculations \citep{Truelove1997, Federrath2011}. 

Lagrangian simulations in the literature have generally been as computationally expedient and offer comparable dynamic range as AMR simulations. However, performing hyper-Lagrangian refinement on a single region of interest can extend the dynamic range even further. For instance, \cite{Hopkins2024a} use \code{GIZMO} to simulate a quasar accretion disk around a super-massive black hole and achieve a dynamic range of $\sim\,13$ orders of magnitude or an effective resolution, $N_{\rm eff} \sim 10^{39}$. The hyper-Lagrangian refinement method involves splitting finite mass fluid particles/cells into daughter particles.  

Splitting fluid particles as a means of further enhancing resolution in Lagrangian simulations has long been a promising avenue. \cite{Kitsionas2002} applied a particle splitting method for SPH on the \cite{Boss1979} test problem and found results comparable to AMR while making considerable computational savings. Similar to \cite{Kitsionas2002}, who place daughter particles in a spherical configuration, authors have experimented with other symmetries \citep{Feldman2007, Chiaki2015}. Several authors have also experimented with placing particles at different separations outright or shuffling them around to reduce the errors \citep[e.g.][]{Lastiwka2005, ReyesLopez2013, Vacondio2013, Gao2022, Nealon2025}. 
Instead of splitting particles on-the-fly (during the simulation), particle splitting can also be used to upsample resolution in a snapshot and using it as an initial condition for a re-simulation. For example, \cite{Bending2020} upsample resolution by up to a factor of 311 by placing particles in concentric shells of increasing radius. This approach has been used in several works such as \cite{Rey-Raposo2015, Rey-Raposo2017, Bending2020, Dobbs2022a, Ali2023} to simulate star forming regions in spiral arms of galaxies.

In \code{GIZMO}, the total domain volume is partitioned among cells\footnote{Throughout this paper, we use the term particles and cells interchangeably when referring to resolution elements in \code{GIZMO}} such that each cell has a volume associated with it (given by \autoref{eq:fractional-volume}) which depends on the smoothing length and number of neighbors (see \autoref{sec:particle-splitting-methods}). The mass, momentum and energy are conserved in a particle splitting operation. However, the new volume per particle is now different since the number density increases. This effectively introduces density perturbations that eventually propagate away and damp out through numerical dissipation, although the rate at which this takes place has not been quantified.  

Concerningly, it is not clear whether these perturbations can seed the growth of structure that would otherwise not exist. It is also not well-understood if the density perturbations exacerbate the sensitivity of star particle formation algorithms to the resolution. These perturbations might therefore lead to the formation of sink particles with different masses (than would have otherwise formed without particle splitting), and bias the inferred initial mass function (IMF), star formation rate (SFR) and feedback yields in star formation calculations using this method. Moreover, we wish to apply the particle refinement technique in zoom-in simulations of Giant Molecular Clouds (GMCs) embedded in galaxies that span several orders of magnitude in spatial scales. The problem becomes readily apparent for such calculations where we do not know what the underlying gas density distribution ought to be. 

Isolated cloud simulations also suffer from this initial condition challenge. The fix so far has been to generate random initial velocities that are sampled from a power spectrum mimicking Burgers (supersonic) turbulence. Since GMCs are chaotic systems, perhaps an ``attractor" solution for the star formation history exists where different random seed trajectories eventually end up. In such a situation, as long as the quantity of interest (e.g. the star formation history) produced in simulations with particle refinement end up in the same attractor, the method can be considered converged. We argue that, equivalently, if the numerical errors introduced by particle refinement (from low to high resolution) are within the inherent random seed variance present in simulations without particle splitting at high resolution, this ``weak convergence" condition is satisfied. A particle splitting run can then be considered to be sampling a trajectory from the space populated by the set of different initial random seeds.

In this paper, we test \code{GIZMO}'s particle splitting algorithm \citep{Angles-Alcazar2021, Franchini2022, Hopkins2024a} and answer the following questions: (a) Does particle splitting seed spurious fragmentation or perhaps suppress fragmentation by being overly diffusive/dissipative? (b) Does particle splitting alter the star formation outcome? 
(c) Does particle splitting generate the right turbulence statistics? 
We use a set of $\sim$ 220 simulations to address these questions. We systematically vary cloud parameters like the mass and radius, feedback physics, initial resolution, and the pace and cadence at which we split particles. In addition to this, we also vary the random seed used to generate the initial turbulent velocity field. We place the differences seen in runs with and without particle splitting in the context of the differences seen in varying the initial random seed at the same resolution.

The rest of the paper is organized as follows: in \autoref{sec:Methods}, we describe our simulations and particle splitting method, and briefly review our sub-grid methods for sink/SSP particles. In \autoref{sec:Results}, we compare the star formation outcome and turbulence statistics in runs where we enhance resolution against runs where we have natively high resolution. 
Lastly, in \autoref{sec:Summary}, we summarize our findings.

\section{Methods and Simulations}
\label{sec:Methods}

\begin{table*}[]
\centering
\renewcommand{\arraystretch}{1.2}
\begin{tabular}{|c|c|c|c|c|c|c|}
\hline
\shortstack{\textbf{Cloud} \\ \textbf{case}}  & \shortstack{\textbf{Resolution} \\ \textbf{($\Delta m_i$) [${\rm M}_\odot$]}}  & \shortstack{\textbf{Star formation} \\ \textbf{algorithm}} & \shortstack{\textbf{Feedback} \\ \textbf{variation}} & \shortstack{\textbf{Random seed} \\ \textbf{variation}} & \shortstack{\textbf{Splitting} \\ \textbf{ends by}} & \textbf{Line-style} \\
\hline

M2e4\_R10 & $ 0.1 $ & STARFORGE & Y & Y & --- & blue solid line \\
M2e4\_R10 & $ 1.6 $ & STARFORGE & Y & N & --- & cyan dotted line \\
M2e4\_R10 & $ 12.8 $ & STARFORGE & Y & N & --- & red dotted line \\
M2e4\_R10 & $ 1.6 \rightarrow 0.1 $ & STARFORGE & Y & N & $0.25\,t_{\rm ff}$ & orange dash-dotted line \\
M2e4\_R10 & $ 12.8 \rightarrow 0.1  $ & STARFORGE & N & N & $0.25\,t_{\rm ff}$ & purple dash-dotted line \\
\hline

M2e5\_R30 & $ 1 $ & STARFORGE & Y & Y & --- & blue solid line \\
M2e5\_R30 & $ 16 $ & STARFORGE & Y & N & --- & cyan dotted line \\
M2e5\_R30 & $ 128$ & STARFORGE & Y & N & --- & red dotted line \\ 

M2e5\_R30 & $ 1 $ & FIRE & Y & Y & --- & green dashed line \\
M2e5\_R30 & $ 16 $  & FIRE & N & N & --- & brown dotted line \\
M2e5\_R30 & $ 128 $  & FIRE & N\footnote{this run and the one above this are run only without feedback. All other runs with N in this column have feedback} & N & --- & moccasin dotted line \\
M2e5\_R30 & $ 16 \rightarrow 1 $ & HYBRID & Y & N & $0.25\,t_{\rm ff}$ & orange dash-dotted line \\
M2e5\_R30 & $ 128 \rightarrow 1$ & HYBRID & Y & N & $0.25\,t_{\rm ff}$ & purple dash-dotted line \\
\hline

M2e7\_R300 & $ 100 $ & FIRE & Y & Y & --- & green dashed line \\
M2e7\_R300 & $ 1600 $ & FIRE & Y & N & --- & brown dotted line \\
M2e7\_R300 & $ 12800 $ & FIRE & Y & N & --- & moccasin dotted line \\
M2e7\_R300 & $ 1600 \rightarrow 100 $ & FIRE & Y & N & $0.25\,t_{\rm ff}$ & orange dash-dotted line \\
\hline \hline
M2e3\_R3 & $ 0.001 $ & STARFORGE & Y\footnote{the runs without feedback have a lower resolution ($\Delta m_i = 0.01$)} & Y\footnote{the code version used for the random seed variation runs is different (taken from \cite{Grudic2023b})} & --- & blue solid line \\
M2e3\_R3 & $ 0.016 \rightarrow 0.001 $   & STARFORGE & N & N & $0.5\,t_{\rm ff}$ & orange dash-dotted line \\
M2e3\_R3 & $ 0.128 \rightarrow 0.001 $  & STARFORGE & N & N & $0.5\,t_{\rm ff}$ & purple dash-dotted line \\
M2e3\_R3 & $ 1.024 \rightarrow 0.001 $  & STARFORGE & N & N & $0.5\,t_{\rm ff}$ & olive dash-dotted line \\
\hline
\end{tabular}
\caption{Summary table for the simulations performed for this study. Column (1): Cloud case (see text for an explanation). Column (2): Resolution ($\Delta m_i$). The right arrow indicates the initial and final resolution in \ps runs. Column (3): The star particle formation algorithm used. Simulations with a HYBRID algorithm have a $\Delta m_i=1 \, {\rm M}_\odot$ and can use either the STARFORGE or FIRE prescriptions based on the cell mass. Column (4): Feedback variation (yes or no). Runs with `Y' have a corresponding run where feedback has been turned off. Column (5): Random seed variation (yes or no). Runs with `Y' have a total of 10 runs with 10 different random seeds. All random seed variation runs are indicated by thinner, high-transparency line-styles compared to the fiducial run. Column (6): The time by which splitting concludes (see text for an explanation). Column (7): Line-style used in the plots in this paper. Random seed variations have a lower opacity value, but the same line-style.}
\label{tab:sim-summary}
\end{table*}

We perform a set of $\approx$ 220 simulations of star forming GMCs, systematically varying (a) cloud parameters like the mass and radius, (b) feedback physics, (c) initial resolution and (d) rate at which we split particles. The simulations evolve the radiation magneto-hydrodynamic (RMHD) equations (including gravity) using a meshless finite mass (MFM) scheme implemented in the \code{GIZMO}\footnote{\href{https://bitbucket.org/phopkins/gizmo-public/src/master/}{https://bitbucket.org/phopkins/gizmo-public/src/master/}} code. 
The physics included in our numerical setup is the same as that described in \cite{Hopkins2024a}. However, we do not include cosmic rays or non-ideal MHD (ambipolar diffusion, Ohmic resistivity and Hall effect) terms and we assume a lower reduced speed of light $\tilde{c} = 10^{-3} c$. 
We refer the reader to \cite{Hopkins2024a} and references therein for the detailed numerical methods, but given the focus of this paper, we briefly review the star formation algorithms below in \autoref{sec:star-formation-methods}.

\subsection{Initial conditions and simulation parameters}
\label{sec:ICs}

All our simulations start from uniform density ($\rho_0$) gas spheres in pressure equilibrium with a warm diffuse medium, with a density $\rho_0/1000$. We consider 4 different cloud cases, corresponding to cloud masses, $M_{\rm cloud}= 2\times10^3, \, 2\times10^4, \, 2\times10^5 \,  {\rm and}\,  2\times10^7 \, \Msun$ with radii $R_{\rm cloud}=$ 3, 10, 30 and 300 pc respectively, such that the surface density is similar ($63.7 \, \Msun\, pc^{-2}$ for the $2\times10^4\, \Msun$ cloud and $70.3\,\Msun\,pc^{-2}$ for all others). We denote a simulation with a cloud mass $2\times10^5\, \Msun$ and radius $30 \, \rm pc$ as the \code{M2e5\_R30} cloud case throughout this paper. The clouds are threaded by a magnetic field (in the z-direction) that contains a tenth of the gravitational binding energy. This corresponds to a field strength of 6.46$\mu G$ for the $2\times10^4\,\Msun$ cloud and 7.18$\mu G$ for all other cloud masses. The clouds are in virial equilibrium with $\alpha_{\rm vir}=2$ (or $E_{\rm turb} = |E_{\rm grav}|$). The velocity dispersion is set by the virial parameter, and can be estimated as $\sigma_v^2 = \alpha_{\rm vir}GM_{\rm cloud}/5R_{\rm cloud}$.

We use different algorithms to follow unresolved fragmentation, based on the gas mass resolution (see \autoref{sec:star-formation-methods}). The STARFORGE sink formation prescription is used for the clouds with $M_{\rm cloud}=$ $2\times10^3$, $2\times10^4$, and $2\times10^5 \, \Msun$. The FIRE SSP particle formation prescription is used for clouds with $M_{\rm cloud}=$ $2\times10^5$ and $2\times10^7 \, \Msun$. 
The initial gas mass resolution (which is fixed for runs where we do not split particles) is $\Delta m = M_{\rm cloud}/2\times 10^5$ for the \code{M2e4\_R10}, \code{M2e5\_R30} and \code{M2e7\_R300} cloud cases. This corresponds to $\Delta m=0.1, 1, 100$ for the 3 cases respectively and allows us to test different star particle formation algorithms. For the \code{M2e3\_R3} cloud case, we use $2\times 10^6$ particles instead, in order to resolve the initial mass function of sink particles. For all cloud cases, we perform runs with and without feedback at the same resolution, except for the \code{M2e3\_R3} case, where we use $2\times 10^5$ particles for the run without feedback (due to computational cost constraints). 

We perform control runs at lower resolutions for all clouds. Simulations with particle splitting start from these lower resolutions and gradually ramp up to the high resolution in less than one free-fall time. We describe our splitting algorithm in detail in \autoref{sec:particle-splitting-methods}. 

At the start of the simulation, particles receive a random velocity kick interpolated from a velocity field generated to mimic the $k^{-2}$ power spectrum seen in supersonic (Burgers) turbulence. Varying the random seed leads to a change in the initial velocity field. We generate initial conditions using the \code{MakeCloud} script\footnote{\href{https://github.com/mikegrudic/MakeCloud}{https://github.com/mikegrudic/MakeCloud}}. All the runs described so far have the same random seed. However, the runs without particle splitting at a high fixed resolution (for each cloud mass) have 9 additional random seed counterparts, making a total of 10 high resolution realizations for a given cloud mass and radius. For the \code{M2e3\_R3} cloud case with feedback, we have 99 additional random seed counterparts, albeit run with an earlier code version\footnote{for a full list of changes between different STARFORGE versions, see \href{https://starforge-tools.readthedocs.io/en/latest/wiki_pages/version_history.html}{https://starforge-tools.readthedocs.io/en/latest/wiki\_pages/version\_history.html}} presented in \cite{Grudic2023b}. 

We typically evolve the cloud for 5 free-fall times. However, not all runs reach $5 \, \tff$, since in several cases, we terminate them early. This is especially true when there are too many star particles orbiting close to each other (more frequently encountered in runs without feedback and/or $M_{\rm cloud}\geq 2\times10^5 \, \Msun$), reducing the timestep and making it prohibitively computationally expensive to evolve further. We summarize all our runs in \autoref{tab:sim-summary}.

\subsection{Particle splitting: when, where and how?}
\label{sec:particle-splitting-methods}

Resolution elements in MFM/MFV methods can either be finite mass or finite volume and have a kernel (smoothing length) associated with them. Each particle $i$ has a fractional volume $\psi_i (\mathbf{x})$ associated with it at a given location $\mathbf{x}$, defined as
\begin{equation}
\label{eq:fractional-volume}
    \psi_i (\mathbf{x}) = \frac{1}{\omega (\mathbf{x})}    W (\mathbf{x} - \mathbf{x}_i, h_i), 
\end{equation}
where $h_i(\mathbf{x})$ is the smoothing length, $W$ is the kernel function (typically cubic spline defined as in Equation H4 in \citealt{Hopkins2015}). The overall normalization $\omega$ is defined as:
\begin{equation}
\label{eq:kernel-normalization}
     \omega (\mathbf{x}) = \sum_j W (\mathbf{x} - \mathbf{x}_j, h_j).
\end{equation}
In SPH schemes, the smoothing length ($h_i$) is often set to enclose a given number of neighboring particles. However, in \code{GIZMO}, the smoothing length (in 3D) is defined based on the smoothed particle number density ($n_i$) as
\begin{equation}
\label{eq:smoothing-length}
    h_{i} = \left(  \frac{3 N_{\rm ngb}}{4 \pi n_i} \right)^{1/3} 
\end{equation}
where $N_{\rm ngb}$ is a constant value (usually set to 32), and $n_i$ is further defined as:
\begin{equation}
\label{eq:number-density}
    n_i = \sum_j W (\mathbf{x}_i - \mathbf{x}_j, h_i).
\end{equation}
The smoothing length is calculated by solving \autoref{eq:smoothing-length} iteratively. The effective volume per particle can then be defined as $V_i = \frac{4 \pi}{3} h_i^3$. The density of a particle in the code is defined as: 
\begin{equation}
\label{eq:density-formula}
    \rho_i = \Delta m_i / V_i = \frac{3 \Delta m_i \, N_{\rm ngb}}{4 \pi \, h_i^3}
\end{equation}

\begin{figure*}
    \includegraphics[width=\textwidth]{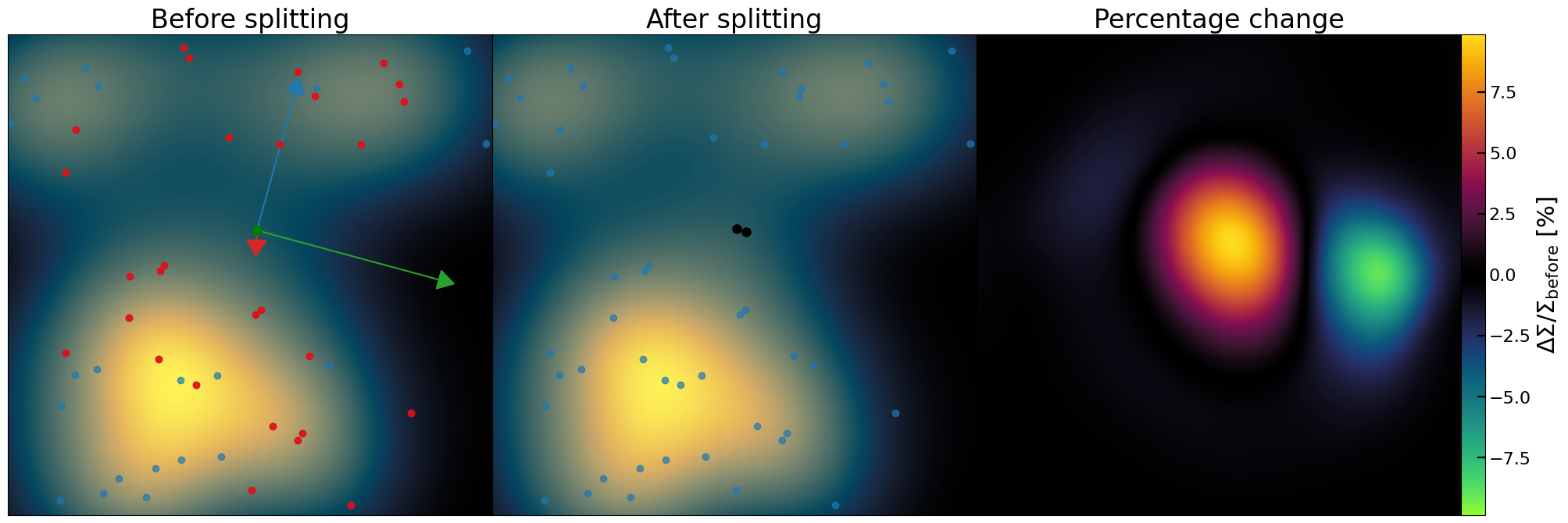}
    \caption{A cartoon illustrating the splitting procedure used in this work.  
    \textit{Left:} The surface density (in arbitrary units) of a random distribution of particles in a 3D box of side length unity.
    Circles denote particle positions. The green circle marks the particle undergoing a split and the red circles denote the particles in its kernel. The arrows correspond to the eigenvectors calculated from the kernel-weighted moment of inertia tensor (\autoref{eq:mom-inertia-tensor}), scaled according to their corresponding eigenvalues. The green arrow corresponds to the largest eigenvalue and determines the direction of the split. \textit{Center:} The surface density distribution after a particle is split. Two new particles with half the original mass (black circles) are placed at a distance $\Delta x$ from the original parent position (found using \autoref{eq:splitting_distance}). \textit{Right:} The percentage change in the surface density maps before and after splitting a particle.}
    \label{fig:splitting_cartoon}
\end{figure*}

\autoref{fig:splitting_cartoon} illustrates our splitting procedure which proceeds as follows. In a splitting operation, a particle is split into 2 daughter particles at a time \citep{Franchini2022}. These daughter particles are assigned half the mass of the parent particle and all parent particle quantities are either copied over to both particles, or divided into two (where appropriate) to ensure mass, momentum and energy are conserved. To decide where to place particles, we solve the eigenvalue problem for a kernel-weighted moment of inertia tensor centered at particle $i$:
\begin{equation}
\label{eq:mom-inertia-tensor}
    \mathcal{T}_i = \sum_j w_{k,j}\begin{bmatrix}
        \Delta x_{ji}^2 & \Delta x_{ji} \Delta y_{ji} & \Delta x_{ji} \Delta z_{ji} \\
         \Delta x_{ji} \Delta y_{ji} & \Delta y_{ji}^2 &  \Delta y_{ji} \Delta z_{ji}\\
         \Delta x_{ji} \Delta z_{ji} & \Delta y_{ji} \Delta z_{ji} & \Delta z_{ji}^2\\ 
    \end{bmatrix},
\end{equation}
where $j$ denotes the sum over the kernel neighbors, $\Delta x_{ji}$, $\Delta y_{ji}$, and $\Delta z_{ji}$ are the distances between particles $j$ and $i$, and $w_{k,j} = W (\mathbf{x}_i - \mathbf{x}_j, h_i)$ is the kernel weight. 
The eigenvector corresponding to the maximum eigenvalue denotes the sparsest direction. 
The new daughter particles are then placed along this axis, equidistant from the parent position. We use this sparsest sampling direction approach to avoid pathological particle configurations and large condition numbers. 
To decide the distance between the daughter particles and the parent particle ($\Delta x_{\rm new}$), we first restrict the distance to the effective particle size in the range $[0.1, 3] h_i/2$, and then limit it further to lie between $[0.1,3]r_{i,\rm near}$ (where $r_{i,\rm near}$ is the distance to the nearest neighbor) before scaling it down yet again. This can be expressed succinctly as:
\begin{equation}
\label{eq:splitting_distance}
\begin{aligned}
\Delta x_a &= \max\left( \Delta x_{\rm eff},\ \frac{0.1 h_i}{2} \right) \\
\Delta x_b &= \min\left( \Delta x_a,\ \frac{3 h_i}{2} \right) \\
\Delta x_c &= \max\left( \Delta x_b,\ 0.1 r_{i, \rm near} \right) \\
\Delta x_d  &= \min\left( \Delta x_c,\ 3 r_{i,\rm near} \right) \\
\Delta x_{\rm new}   &= 0.39685 \cdot \Delta x_d
\end{aligned}
\end{equation}
where $\Delta x_{\rm eff} = \left( \frac{4\pi}{3} \right)^{1/3} \frac{h_i}{N_{i, {\rm ngb}}}$, $h_i$ is the smoothing length of the particle, $N_{i, {\rm ngb}}$ is the number of nearest neighbors of the parent particle $i$.

The daughter particles initially inherit the smoothing length of the parent particle. In the next timestep, the number density of all cells surrounding the daughter particles as well as that of the daughter particles themselves also changes due to a change in the number of neighbors. Particle splitting is performed gradually (in time and space) by waiting 30-300 timesteps before splitting again. The details of the splitting operation used in this work largely follows from earlier works by \cite{Angles-Alcazar2021, Franchini2022, Hopkins2024a, Hopkins2024b, Hopkins2024c}. However, we follow a different method to decide when and where to split particles which we discuss next.

We wish to eventually apply the particle splitting method in large dynamic range zoom-in simulations that span spatial scales across several orders of magnitude. Given that the density perturbations require additional time to damp out, it would be prudent to implement a refinement scheme that consists of a relaxation stage between splitting stages, so that the turbulent cascade and density structure has time to develop. Note that this is applied in addition to waiting a certain number of timesteps for each particle.

We therefore perform refinement using a time-dependent multi-zone function for our target resolution. We use a series of time periods where we either split particles or let them relax. In a relaxation phase, the target resolution function is frozen in. During this phase however, individual particles can still undergo splitting to reach their target resolution since each particle has its own timestep and may not have been able to undergo a split in the splitting phase. We define a time period $\Delta t_{\rm split}$, that denotes the duration of the splitting phase. The relaxation phase lasts for $2\Delta t_{\rm split}$ after each splitting phase. We use 1 splitting and 1 relaxation phases (one refinement cycle) for every order of magnitude change in the target resolution. 
Thus, in our simulations where we increase resolution by a factor of 128, we use 2 refinement cycles. We set the timescale for this process to be $\Delta t_{\rm split} = f_{\rm split} \tff/6$, where $f_{\rm split}=0.25$ or 0.5. If $f_{\rm split}=0.25$, we are done with the 2 refinement cycles by $0.25 \, \tff$.

For each refinement zone \( i \) with inner and outer radii \( r_{\text{in}, i} \) and \( r_{\text{out}, i} \), the target mass resolution is defined as:
\begin{equation}    
\label{eq:splitting_function}
\ln (m_{\rm{target}})(r, t) = 
\begin{cases}
\ln (m_{\text{init}, i}), & r \geq r_{\text{out}, i} \\
\alpha_i(t) \ln (r/r_{\rm in}) + \beta_i(t), & r_{\text{in}, i} \leq r < r_{\text{out}, i} \\
\beta_i(t), & r < r_{\text{in}, i}
\end{cases}
\end{equation}

The coefficients \( \alpha_i(t) \) and \( \beta_i(t) \) depend on the phase. While splitting:
\begin{equation}
\begin{aligned}    
\beta_i(t) &= \ln m_{\text{init}, i} - \ln\left( \dfrac{m_{\text{init}, i}}{m_{\text{final}, i}} \right) \cdot \dfrac{t - t_{\text{start}}}{\Delta t}   \\ 
\alpha_i(t) &= \dfrac{\ln m_{\text{init}, i} - \beta_i(t)}{\ln \left( \dfrac{r_{\text{out}, i}}{r_{\text{in}, i}} \right)},
\end{aligned}
\end{equation}
and while relaxing:
\begin{equation}
\begin{aligned}
\beta_i(t) &= \ln m_{\text{final}, i}, \\
\alpha_i(t) &= \dfrac{\ln m_{\text{init}, i} - \ln m_{\text{final}, i}}{\ln \left( \dfrac{r_{\text{out}, i}}{r_{\text{in}, i}} \right)}.
\end{aligned}
\end{equation}
Here $i\in\{1,2\}$, $m_{{\rm init}, i}$ and $m_{{\rm final}, i}$ denote the initial and final mass resolution for a given zone ($m_{{\rm final},i}=m_{{\rm init}, i+1}$), $t$ is the total time since the start of the simulation and $t_{\rm start}=\{0,3\Delta t_{\rm split}\}$ denotes the start of the splitting phases. All particles within the inner edge of a refinement zone ($r<r_{{\rm in},i}$) or beyond the outer edge ($r\geq r_{{\rm out},i}$) have the same mass resolution. 

For particle splitting runs, our starting resolution can be 2 or 3 orders of magnitude coarser than the final resolution.
We split particles throughout the cloud (i.e., $r_{\rm out,1}>r_{\rm in,1}>r_{\rm out,2}>r_{\rm in,2}=1.75 R_{\rm cloud}$). We gradually split particles until we achieve the same resolution as the natively higher resolution runs after $4\Delta t_{\rm split}$. We note here that particles can sometimes take longer to achieve their final resolution since we typically wait 300 time-steps before a daughter particle can be split again. Unlike typical star formation simulations, the refinement strategy we use here does not maintain a certain resolution of the Jeans length (or mass). However, in practice, our refinement ends by $\sim 0.25$ or $0.5\, \tff$, well before any star particles form, which ensures that the \cite{Truelove1997} criterion is always satisfied.

\subsection{Star formation}
\label{sec:star-formation-methods}

We use two different prescriptions for star formation, based on the gas mass resolution. We refer to (unresolved) collapsed objects as either single-age stellar population (SSP; non-accreting) or individual star (sink; accreting) particles, depending on the prescription used. We refer the reader to \cite{Hopkins2023, Hopkins2024a} for more details on the FIRE-3 SSP particle formation algorithm, but briefly review it here for completeness. The FIRE SSP particle formation prescription is used for gas particles with mass $\Delta m_i>1 \Msun$. Gas is converted into SSP particles, when certain criteria are satisfied. Eligible cells are assigned a SFR as 
\begin{equation}
    \dot{\rho}_{*,i} = \epsff \frac{\rho_i}{t_{*,i}}
\end{equation}
where the timescale for star formation, $t_{*,i} = {\rm min}[t_{{\rm ff}, i}, |\nabla\cdot \mathbf{v}|_i^{-1}]$ if $\nabla\cdot \mathbf{v}<0$ and $t_* = \tff$ otherwise, $\rho_i$ is the density of the gas cell, $\epsff=1$ is the star formation efficiency per free-fall time. 
To qualify for star formation, a gas cell has to be (a) Jeans unstable ($m_J < {\rm max[100 \, M_\odot}, \Delta m_i]$, where $m_J$ is the thermal Jeans mass and $\Delta m_i$ is the cell mass), (b) part of a converging flow ($\nabla\cdot \mathbf{v}<0$), and (c) sub-virial ($\alpha_i<1$). The virial ratio is calculated as $\alpha_i = (||\nabla\otimes \mathbf{v}||_i^2 + 2[(c_{s, i}^2 + v_{A,i}^2)/\Delta x_{{\rm eff},i}])/(8 \pi G \rho_i)$, where $|| \, ||$ denotes the Frobenius norm, $c_s$ and $v_A$ are the sound and Alfv{\'e}n speeds respectively, and the remaining quantities carry their usual meaning.

SSP and sink particles evolve according to their respective algorithms. SSP particles evolve according to fits to the STARBURST99 \citep{Leitherer2014} model as detailed in \citet{Hopkins2023}. They eject mass, momentum, energy and metals via supernovae (core-collapse and Type Ia), stellar winds (from O/B and AGB stars). Photon momentum fluxes and radiative heating are determined via stellar population spectra calibrated for a \cite{Kroupa2001} IMF \citep{Hopkins2020}. Unlike \cite{Hopkins2024a}, we do not inject cosmic rays in these runs. We quantize the feedback rates with the intent of mimicking the discreteness of feedback rates as described in \citet{Su2018, Grudic2019b}. The SSP particles carry a number of O-stars drawn from a Poisson distribution. This is then used to scale the IMF-averaged bolometric and ionizing luminosity, and the mass-loss rates from young star stellar winds. It also sets an upper limit on the number of supernovae (SNe) from a given particle.

The STARFORGE sink particle algorithm is described in \cite{Grudic2021}. We use this algorithm when the gas particle mass is $\Delta m_i\leq1 \Msun$. Gas cells have a stricter set of eligibility criteria to convert to sink particles in this case, given as:
(a) cells must exceed a density threshold defined as: 
\begin{equation}
    \rho_{\rm threshold} = \rho_J = f_{J, {\rm max}}^6 \frac{\pi^3 c_s^3}{G^3 \Delta m^2},
\end{equation}
where $f_{J, \rm max}=1/2$ and other symbols carry their usual meaning, (b) cells must be at a density maximum (no sinks should overlap with the kernel radii), (c) cells must be part of a converging flow, and (d) sub-virial. In addition to this, (e) the tidal tensor defined at the position of the cell must be fully compressive (possess negative eigenvalues) and (f) the free-fall time of the cell must be smaller than the timescale for approaching or orbiting around another sink particle, i.e., 
\begin{equation}
    t_{{\rm ff}, i} = \sqrt{\frac{3 \pi}{32 G \rho_i}} < {\rm min} \left[ \frac{\sqrt{r_{ij}^2 + \epsilon_{*,j}^2}}{v_{ij}}, \sqrt{\frac{(r^2_{ij}+\epsilon_{*,j}^2)^{3/2}}{G (\Delta m_i + m_j)}} \right]
\end{equation}
where the subscript $j$ refers to other (existing) sink particles, $\epsilon_{*,j}$ is the Plummer-equivalent sink softening radius, $r_{ij}$ and $v_{ij}$ are the separation and relative velocities of the gas cell $i$ and sink $j$.

Sink particles can accrete gas based on certain criteria. To be eligible for accretion, a gas cell must: (a) be within the sink particle's radius $R_{\rm sink}$, (b) bound to the sink particle, (c) possess less angular momentum than a circular Keplerian orbit around the sink particle, (d) have a volume less than the volume within $R_{\rm sink}$. If more than one sink particle competes for the same gas cell, the gas is accreted by the sink with the shorter mutual dynamical time $t_{\rm dyn} = \sqrt{r_{ij}^3/G(\Delta m_i + m_j)}$. Sink particles can also merge (although this is very rare) if their binary semi-major axis is less than $R_{\rm sink}$ and the masses are within a factor of 10. 

Sink particles evolve along protostellar and main-sequence tracks as described in \cite{Offner2009} and \cite{Grudic2021}. They produce bipolar jets and stellar winds, and emit radiation as a blackbody with an effective temperature set by the luminosity and radius as determined by the stellar evolution track. At the end of their lives, stars with $M_{\rm ZAMS} > 8 M_\odot$ explode as SNe by injecting their remaining mass and metals into the simulation domain as resolved ejecta, with a kinetic energy of $10^{51} \rm erg$.

\section{Results}
\label{sec:Results}

\begin{figure*}
    \centering
    \includegraphics[width=\textwidth]{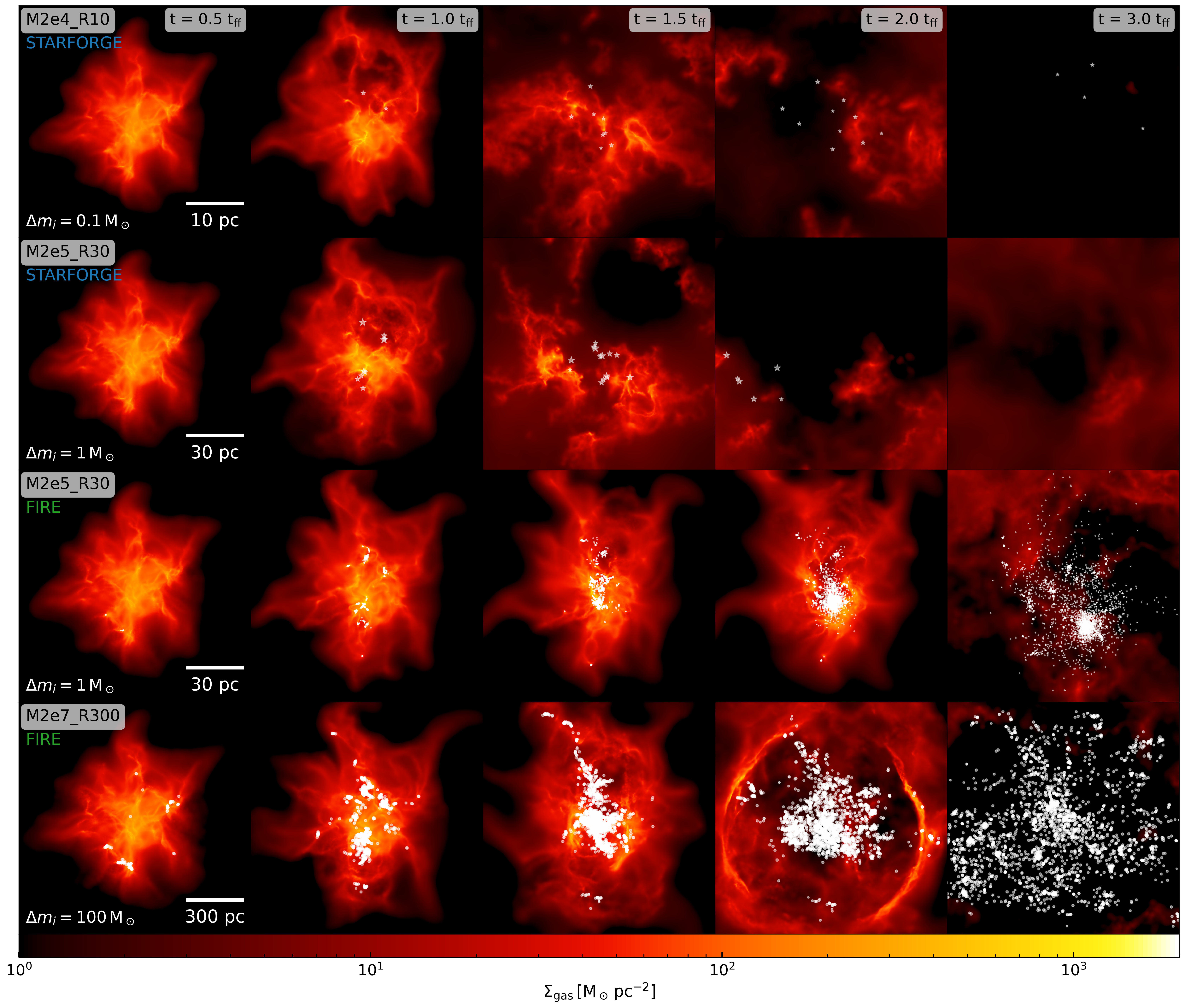}
    \caption{The surface density maps of a subset of our runs without particle splitting, at 5 different times ($0.5 \tff, 1 \tff, 1.5 \tff, 2 \tff$ and $3 \tff$). The white stars (top two rows) indicate STARFORGE sink particles, whereas the while circles indicate FIRE SSP particles. The cloud mass and radius are indicated in the text box at the top left of each row. Indicated below that is the sub-grid star particle formation model used for each row. In the bottom-left, we highlight the gas mass resolution in each of the rows. All of the simulations shown here are run at uniform mass resolution, with the same random seed used to generate the initial turbulent velocity field.}
    \label{fig:surf_dens1}
\end{figure*}

\begin{figure*}
    \centering
    \includegraphics[width=\textwidth]{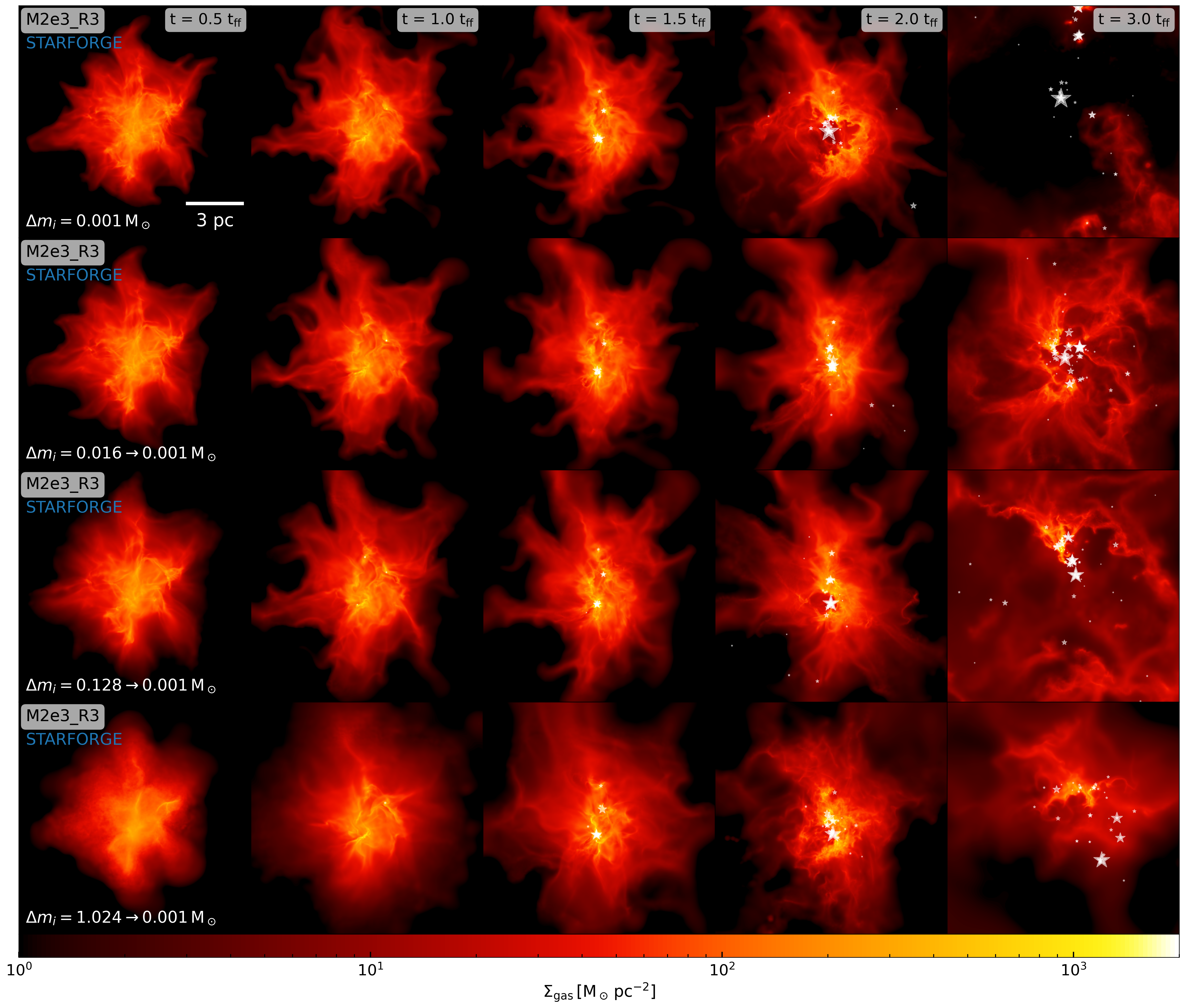}
    \caption{Same as \autoref{fig:surf_dens1}, but comparing runs with (bottom 3 rows) and without (top row) particle splitting. At early times, there are slight differences between the evolution of the cloud at different resolutions. By $1.5\, \tff$, star particles in the \ps runs form in roughly the same location as the natively high resolution run. However, the masses of the stars can be different (as seen in the size of the markers) and thus the feedback from the stars affects the cloud evolution differently at late times in the various runs.}
    \label{fig:surf_dens2}
\end{figure*}

We briefly discuss our expectations first so it is easier to interpret our results. We generally expect lower resolution runs to have a smoother density field than the same run at higher resolution because of missing power in the high-$k$ modes that is not resolved. Since splitting particles is an inherently noisy process, high-$k$ power is added in an uncontrolled manner -- indeed this is why we wait $30-300$ timesteps before splitting again. We therefore do not know whether splitting will tend to produce more, smaller, and earlier star particles, or vice versa. These differences can lead to major downstream consequences especially when feedback is active. 

Detailed convergence to the natively high-resolution result is impossible, especially in the presence of feedback. However, we can seek a ``weak convergence", in which the systematic changes caused by splitting from a coarser-resolution run are nevertheless small in comparison to the typical differences between simulations that are identical except for the random seed. We therefore define a metric for this ``weak convergence" for a given quantity $y(x)$ (star formation history, density distribution, etc), as the ratio of the areas $A_{\Delta m}$ and $A_{1\sigma}$ or $A_{\rm range}$. We define these as:
\begin{equation}
\label{eq:A-dm}
    A_{\Delta m} = \int_{x_{\rm min}}^{x_{\rm max}} | y_1 (x) - y_2 (x)| \, dx,
\end{equation}
where $y_1(x)$ is the natively high resolution run and $y_2(x)$ is the particle splitting run which reaches that same final resolution; 
\begin{equation}
\label{eq:A-sigma}
    A_{1\sigma} = \int_{x_{\rm min}}^{x_{\rm max}} [s_{{84}}(x) - s_{16}(x)] \, dx,
\end{equation}
where $s_{16}(x)$ and $s_{{84}}(x)$ are the 16th and 84th percentiles across the random seed ensemble at each $x$. 
\begin{equation}
\label{eq:A-range}
    A_{\rm range} = \int_{x_{\rm min}}^{x_{\rm max}} [\max_i \, s_i(x) - \min_i s_i(x)] \, dx,
\end{equation}
where $i$ denotes a random seed run. We calculate these areas in the log-log plane and use a fixed integration domain $[x_{\rm min}, x_{\rm max}]$ and $[y_{\rm min}, y_{\rm max}]$. When curves are only defined for some range $[a,b]$, where $a<b$, $a>x_{\rm min}$ and $b<x_{\rm max}$, we adopt two different approaches. For quantities where $x$ denotes time, we extrapolate the curve between $[x_{\rm min}, a]$ and $[b,x_{\rm max}]$ by keeping the curve fixed at $y(x_a)$ for $x<a$ and $y(x_b)$ for $x>b$ respectively. For all other quantities, we use a floor at $y_{\rm min}$.   
We calculate the $1\sigma$ range by discretizing $x$ and calculating the 16$^{\rm th}$ and 84$^{\rm th}$ percentile at each $x$ value. We adopt the same procedure for finding the $\underset{i}{\rm min} \, s_i(x)$ and $\underset{i}{\rm max} \, s_i(x)$. 
We compare $A_{\Delta m}$ with the area bounded by the $16^{\rm th}-84^{\rm th}$ percentile band ($A_{\rm 1\sigma}$) and the entire range ($A_{\rm range}$) of random seed runs at natively high resolution. Our convergence condition then translates to the ratio, $A_{\Delta m}/A_{\rm 1\sigma}<1$ or $A_{\Delta m}/A_{\rm range}<1$. We do not report this ratio for the \code{M2e5\_R30} cloud with feedback, since the particle splitting runs can form both sinks and SSPs, which renders the comparison unfair.

We begin our analysis by examining surface density maps of a subset of our simulations at different times. 
In \autoref{fig:surf_dens1}, we show runs with feedback for the \code{M2e4\_R10}, \code{M2e5\_R30} and \code{M2e7\_R300} clouds. 
The top-middle row shows the \code{M2e5\_R30} with the STARFORGE sink particle formation prescription, whereas the bottom-middle row shows the same cloud with FIRE SSP formation prescription. The cloud disrupts soonest at $\sim1.5 \, \tff$ when sink particles are used. This is because these sink particles start off being very massive given the resolution ($\Delta m_i= 1\,\Msun$) and thus have very energetic feedback that disrupts the cloud sooner than it would otherwise be. This highlights the sensitive dependence of the star formation outcome on the resolution when we use the sink particle prescription. Star formation is less sensitive to the resolution when the FIRE SSP particle formation prescription is used, as we shall see in \autoref{sec:SFH}. 

In \autoref{fig:surf_dens2} we show variations in \code{M2e3\_R3} cloud runs with feedback. In the top row, we show the surface density evolution of the run at natively high resolution ($\Delta m_i= 0.001\,\Msun$) without particle splitting. The \ps runs at progressively lower initial resolution ($\Delta m_i= 0.016, 0.128, 1.024 \,\Msun$) are shown in the middle and bottom rows. The density field at $t=0.5 \, \tff$ in the \ps runs appears smoother with worsening initial resolution. This is because the initial velocity field is interpolated onto a smaller set of particles in lower resolution runs, naturally smoothing out the initial velocity field and inhibiting shocks that would sharpen the density field. By $t=1\,\tff$, the differences appear minimal in the $\Delta m_i= 0.016 \rightarrow 0.001 \,\Msun$ and $\Delta m_i= 0.128 \rightarrow 0.001 \,\Msun$ runs when compared to the $\Delta m_i= 0.001 \,\Msun$ run, as more structure develops. The sink particles form roughly in the same locations in all the runs regardless of starting resolution. However, there are slight differences in the masses of sink particles which ultimately lead to slightly different cloud destruction timescales (as seen at $t=3 \, \tff$) because of the sensitive dependence of the feedback on the (massive) sink masses.

In the following subsections, we study (a) the star formation histories (SFHs), (b) the number of stars, (c) the IMF and the light-to-mass ratios, (d) the density probability distribution functions (PDFs) and the density power spectra, and (e) the velocity power spectra for our runs. 

\vspace{20pt}
\subsection{Star formation history}
\label{sec:SFH}

\begin{figure*}
    \centering
    \includegraphics[width=\textwidth]{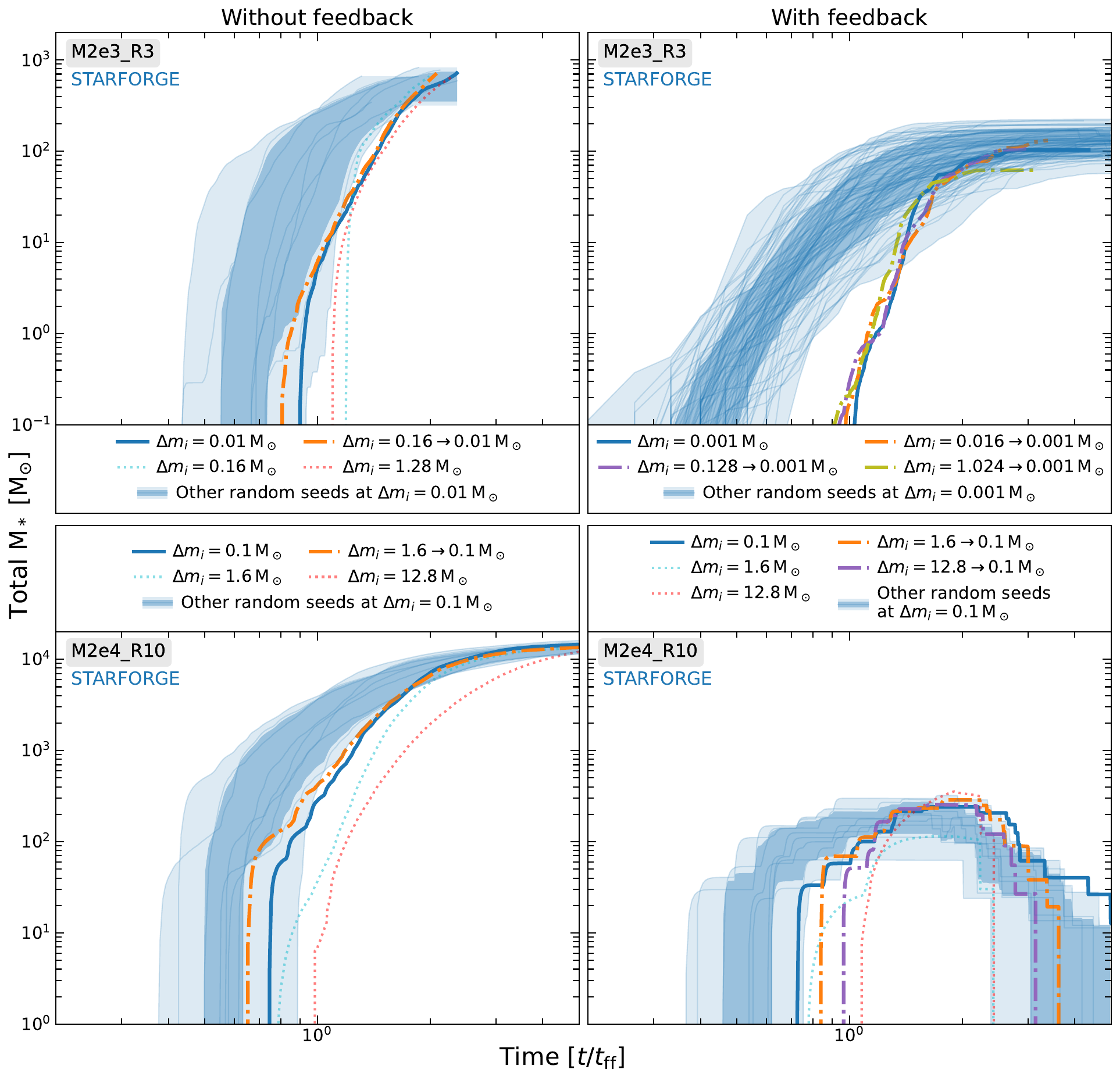}
    \caption[]{The total stellar mass in the simulation as a function of time (in units of $\tff$) for our \code{M2e3\_R3} and \code{M2e4\_R10} runs which without (\textit{left column}) and with (\textit{right column}) feedback. The cloud mass and radius are indicated in the text box at the top left along with the sub-grid model used for each panel. Different lines indicate \ps runs (orange dash-dotted lines), runs with natively high resolution (thick solid blue lines), and runs where we varied the random seed for the initial turbulent velocity field (thin solid blue lines). The shaded bands denote the $16^{\rm th} - 84^{\rm th}$ percentile range (dark) and the entire range (light) for the random seed runs. 
    The 99 random seed runs (thin blue solid lines) in the \code{M2e3\_R3} cloud with feedback are taken from \cite{Grudic2023b} and used an earlier code version (see text for details about the version changes). The total stellar mass formed in \ps runs is close to that in the corresponding natively high resolution runs, especially when compared to the differences seen by varying the initial turbulent random velocity field at high resolution.}
    \label{fig:sfh1}
\end{figure*}

\begin{figure*}
    \centering
    \includegraphics[width=\textwidth]{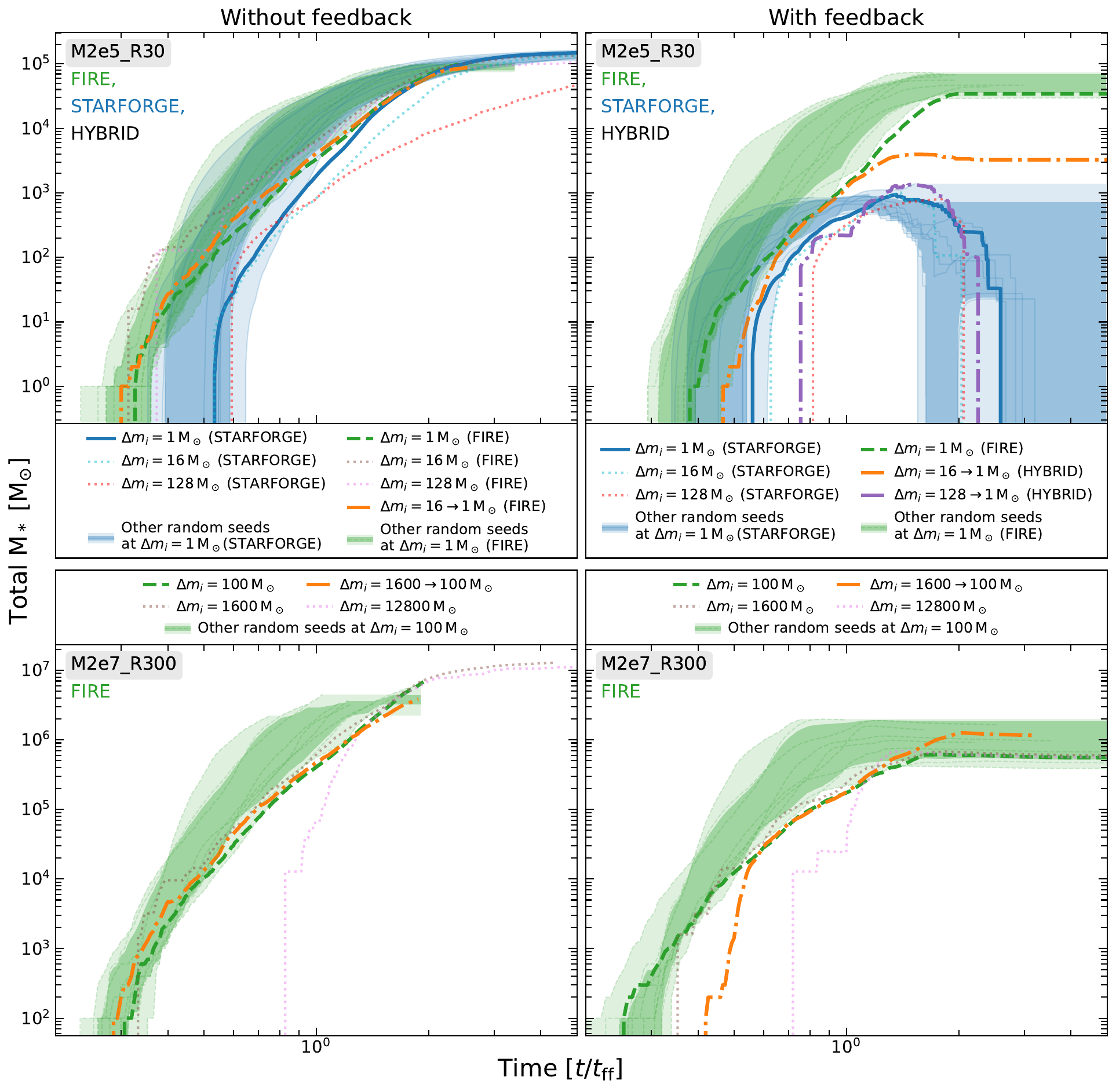}
    \caption[]{Same as \autoref{fig:sfh1}, but for the \code{M2e5\_R30} and \code{M2e7\_R300} runs. The total stellar mass formed in \ps runs is again close to that in the corresponding natively high resolution runs. The differences caused by \ps are smaller than the inherent scatter present when varying the initial random seed for turbulence.}
    \label{fig:sfh2}
\end{figure*}

To understand how our particle splitting affects gravitational collapse into sink and SSP particles, we plot the total mass in star particles as a function of time in \autoref{fig:sfh1} for the \code{M2e3\_R3} and \code{M2e4\_R10} clouds with (right column) and without (left column) feedback. For the \code{M2e3\_R3} cloud without feedback, we find that the \ps run forms stars earlier than the natively high resolution ($\Delta m_i = 0.01 \, \Msun$) run, while also forming a more massive sink particle initially. Both these runs, as well as several of the random seed runs also converge to the same total mass even if the first sink's formation has a large scatter. This is not surprising since the cloud ends up converting $\sim 50\%$ of its mass into stars in the absence of feedback. 
The same trends hold true for the \code{M2e4\_R10} cloud's case without feedback.

For the \code{M2e4\_R10} cloud with feedback, the \ps runs start forming sinks later than the natively high resolution run without particle splitting. While there is still a large variation in when the first star forms, the final (maximum) stellar mass achieved is nevertheless the same across different random seed runs. Runs with \ps also achieve the same maximum stellar mass. The total stellar mass decreases at late times because stars are removed from the simulation when they go supernova.

We note that the runs described above do not have sufficient gas mass resolution to resolve sink formation and growth with enough fidelity. The \code{M2e3\_R3} cloud with feedback has 10 times higher resolution. 
Here we find that the differences between the runs with and without particle splitting reduce significantly for the first sink formation time as well as the overall SFH. We also show runs where we vary the initial random seed for the turbulence to demonstrate the diverse range of SFHs we get. 

\begin{table}
\centering
\setlength{\tabcolsep}{2pt}
\begin{tabular}{cccc}
\toprule
Cloud & $\Delta m_i$ & $A_{\Delta m}/A_{\rm 1\sigma}$ & $A_{\Delta m}/A_{\rm range}$ \\ \hline
\multicolumn{4}{c}{\textbf{With feedback}} \\ \hline
\texttt{M2e3\_R3} & $1.024 \rightarrow 0.001 \, {\rm M}_\odot$ & 0.20 & 0.08 \\
 & $0.128 \rightarrow 0.001 \, {\rm M}_\odot$ & 0.08 & 0.03 \\
 & $0.016 \rightarrow 0.001 \, {\rm M}_\odot$ & 0.12 & 0.05 \\
\texttt{M2e4\_R10} & $12.8 \rightarrow 0.1 \, {\rm M}_\odot$ & 0.67 & 0.38 \\
 & $1.6 \rightarrow 0.1 \, {\rm M}_\odot$ & 0.42 & 0.24 \\
\texttt{M2e5\_R30} & \multicolumn{3}{c}{\ldots} \\
\texttt{M2e7\_R300} & $1600 \rightarrow 100 \, {\rm M}_\odot$ & 0.35 & 0.25 \\ \hline
\multicolumn{4}{c}{\textbf{Without feedback}} \\ \hline
\texttt{M2e3\_R3} & $0.16 \rightarrow 0.01 \, {\rm M}_\odot$ & 0.26 & 0.15 \\
\texttt{M2e4\_R10} & $1.6 \rightarrow 0.1 \, {\rm M}_\odot$ & 0.24 & 0.13 \\
\texttt{M2e5\_R30} & $16 \rightarrow 1 \, {\rm M}_\odot$ & 0.16 & 0.10 \\
\texttt{M2e7\_R300} & $1600 \rightarrow 100 \, {\rm M}_\odot$ & 0.34 & 0.20 \\ \hline
\end{tabular}
\caption{Summary statistics for deviations in the SFH while splitting particles (\autoref{fig:sfh1} and \autoref{fig:sfh2}). Columns (1-2): Cloud case and resolution. Column (3-4): The ratio of area between the run with and without particle splitting ($A_{\Delta m}$; \autoref{eq:A-dm}) to the seed-to-seed scatter area, measured against the $16^{\rm th}-84^{\rm th}$ percentile band ($A_{\rm 1\sigma}$; \autoref{eq:A-sigma}) and the full range ($A_{\rm range}$; \autoref{eq:A-range}). All the ratios are less than unity. Thus the differences caused by particle splitting are less than the scatter due to our choice of initial conditions.}
\label{tab:sfh_area}
\end{table}

In \autoref{fig:sfh2}, we consider the \code{M2e5\_R30} and \code{M2e7\_R300} cloud cases. For the \code{M2e5\_R30} cloud, we run two simulations at our fiducial high resolution ($\Delta m_i = 1 \, \Msun$), one where we use the FIRE SSP particle formation algorithm (green dashed line), and the other where we use STARFORGE's sink particle algorithm (blue solid line). The particle splitting run (dash-dotted orange line) can form either SSP particles (if $\Delta m_i > 1 \, \Msun$) or sink particles (if $\Delta m_i \leq 1 \, \Msun$), based on the parent gas particle's mass. In the case without feedback, we find that the particle splitting run's SFH is virtually identical to the high resolution run with SSP particles. This trend continues for the \code{M2e7\_R300} cloud without feedback, the particle splitting run follows the natively high resolution run ($\Delta m_i = 100 \, \Msun$) closely. 

When feedback is enabled, both the \code{M2e5\_R30} and the \code{M2e7\_R300} clouds start forming SSPs later when we increase resolution through particle splitting. For the  \code{M2e5\_R30} cloud, in the run where we split particles ($\Delta m_i= 16 \rightarrow 1 \,\Msun$, orange dash-dotted line), SSPs form during the splitting phase of the run and sinks form when the final gas resolution is reached. As a result of this, when sinks go SNe, star formation is inhibited and the SFH at late times looks different from the run at natively high resolution with SSP formation (green dashed line). The particle splitting run with a lower starting resolution ($\Delta m_i= 128 \rightarrow 1 \,\Msun$, purple dash-dotted line) only forms sinks once the final gas mass resolution is achieved. This is a result of the density field being smoothed out (more than in the case of the $\Delta m_i= 16 \rightarrow 1 \,\Msun$ run), which prevents collapse before the simulation has had time to reach its final resolution, at which point it is eligible to form individual sink particles. Hence this run looks similar to the natively high resolution run with sink formation (blue solid line).

We summarize our results from \autoref{fig:sfh1} and \autoref{fig:sfh2} in \autoref{tab:sfh_area}. In all cases, the ratios $A_{\Delta m}/A_{\rm 1\sigma}$ and $A_{\Delta m}/A_{\rm range}$ are less than unity. 
We emphasize here that in all cases, the differences in the \ps runs and the runs at natively high resolution without particle splitting are lesser than the inherent scatter in the SFH present from varying the initial random seed for the turbulence. Effectively, particle splitting introduces errors in the SFH that are less than the variance introduced by the arbitrary choice of initial turbulent phases.

\subsection{Number of stars formed}
\label{sec:num-stars}

\begin{figure*}
    \centering
    \includegraphics[width=\textwidth]{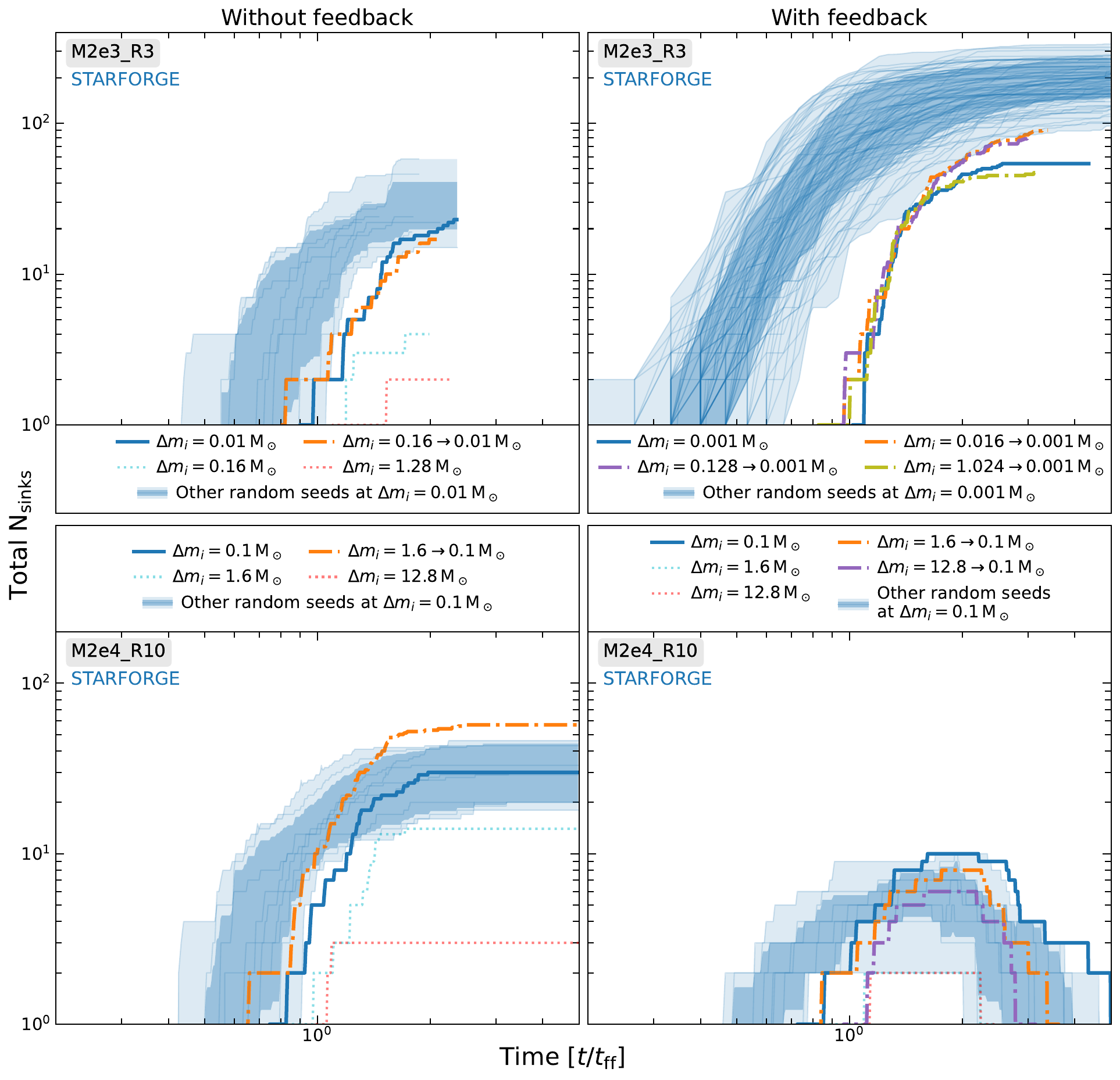}
    \caption{The total number of star (sink) particles in the simulation as a function of time (in units of $\tff$) for our \code{M2e3\_R3} and \code{M2e4\_R10} runs which without (\textit{left column}) and with (\textit{right column}) feedback. The number of sinks formed in \ps runs is within a factor of $\sim 2$ compared to the corresponding natively high resolution runs. This is less than the scatter seen by varying the random seed for the initial turbulent velocity field.}
    \label{fig:num_stars1}
\end{figure*}

\begin{figure*}
    \centering
    \includegraphics[width=\textwidth]{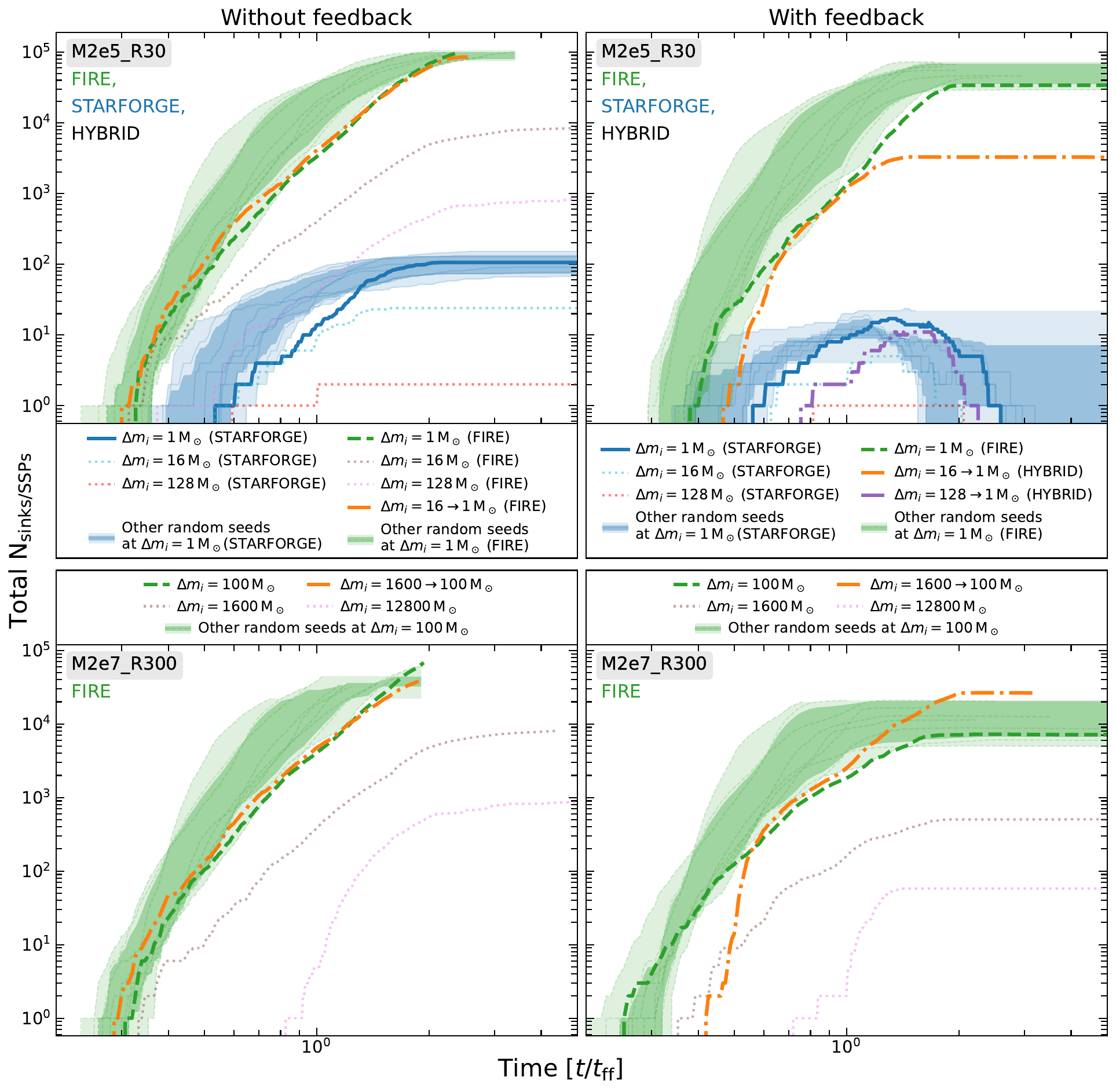}
    \caption{The total number of star (SSP/sink) particles in the simulation as a function of time (in units of $\tff$) for our \code{M2e5\_R30} and \code{M2e7\_R300} runs which without (\textit{left column}) and with (\textit{right column}) feedback. The number of SSPs formed in \ps runs follows the number of SSPs formed in the corresponding natively high resolution runs closely. For the \code{M2e5\_R30} cloud with feedback, both STARFORGE sinks and FIRE SSPs can form (see text for an explanation). Nevertheless, the differences caused by \ps are smaller than the random seed scatter.}
    \label{fig:num_stars2}
\end{figure*}

We show in \autoref{sec:truelove} that we do not see signatures of artificial fragmentation when we perform particle refinement in an idealized isothermal collapsing cloud setup \citep{Boss1979}. However, in a turbulent cloud setup, fluctuations in the density field caused by particle refinement can compound and couple non-linearly with additional physics (e.g. feedback or cooling), such that it leads to differences in the numbers and mass spectra of the star particles that form. While an increase or decrease in the number of sinks/SSPs in a turbulent cloud setup can be considered to be artificial fragmentation or suppression, we make a distinction between the two setups here. Given that our particle refinement scheme does not produce artificial fragmentation in the traditional test \citep{Boss1979, Truelove1997}, we argue that for a turbulent cloud setup, a stricter criteria holds. If the differences in the number of star particles formed between the run with particle refinement and the run at natively high resolution are larger than the seed-to-seed scatter, then the refinement scheme produces artificial fragmentation or suppression. In this section, we study the effect of particle splitting on number and the average masses of star particles in our turbulent cloud setup.

\autoref{fig:num_stars1} shows that the number of sink particles formed in \code{M2e3\_R3} cloud without feedback is similar in both the \ps run (orange dash-dotted line) as well as the natively high resolution run (blue solid line). Even when feedback is enabled, the number of sink particles formed in the \ps runs is within a factor of 2 compared to that in the natively high resolution run.
The number of sink particles formed in \code{M2e4\_R10} cloud with feedback is similar in both the \ps run as well as the natively high resolution run. 
The \code{M2e4\_R10} cloud without feedback forms more sink particles in the runs with particle splitting than in the run at natively high resolution, however, this difference is still within the random seed scatter since $A_{\Delta m}/A_{\rm 1\sigma}$ and $A_{\Delta m}/A_{\rm range}$ are both less than unity (see \autoref{tab:nsinks_area}). Therefore, this does not constitute a signature of artificial fragmentation.

The \code{M2e5\_R30} and \code{M2e7\_R300} clouds without feedback also form the same amount of FIRE SSPs, regardless of whether we split particles or not, as seen in \autoref{fig:num_stars2}. We note that in the \code{M2e5\_R30} cloud, even though the final total stellar mass is the same in all the runs regardless of whether the FIRE SSP or STARFORGE sink particle formation algorithm is used, the numbers of star particles formed are vastly different. STARFORGE sink particles can grow in mass by accreting, are fewer in number and thus very massive on average as a result.

\begin{table}
\centering
\setlength{\tabcolsep}{2pt}
\begin{tabular}{cccc}
\toprule
Cloud & $\Delta m_i$ & $A_{\Delta m}/A_{\rm 1\sigma}$ & $A_{\Delta m}/A_{\rm range}$ \\ \hline
\multicolumn{4}{c}{\textbf{With feedback}} \\ \hline
\texttt{M2e3\_R3} & $1.024 \rightarrow 0.001 \, {\rm M}_\odot$ & 0.08 & 0.04 \\
 & $0.128 \rightarrow 0.001 \, {\rm M}_\odot$ & 0.18 & 0.08 \\
 & $0.016 \rightarrow 0.001 \, {\rm M}_\odot$ & 0.19 & 0.09 \\
\texttt{M2e4\_R10} & $12.8 \rightarrow 0.1 \, {\rm M}_\odot$ & 0.80 & 0.44 \\
 & $1.6 \rightarrow 0.1 \, {\rm M}_\odot$ & 0.46 & 0.25 \\
\texttt{M2e5\_R30} & \multicolumn{3}{c}{\ldots} \\
\texttt{M2e7\_R300} & $1600 \rightarrow 100 \, {\rm M}_\odot$ & 1.24 & 0.69 \\ \hline
\multicolumn{4}{c}{\textbf{Without feedback}} \\ \hline
\texttt{M2e3\_R3} & $0.16 \rightarrow 0.01 \, {\rm M}_\odot$ & 0.35 & 0.20 \\
\texttt{M2e4\_R10} & $1.6 \rightarrow 0.1 \, {\rm M}_\odot$ & 0.55 & 0.38 \\
\texttt{M2e5\_R30} & $16 \rightarrow 1 \, {\rm M}_\odot$ & 0.19 & 0.12 \\
\texttt{M2e7\_R300} & $1600 \rightarrow 100 \, {\rm M}_\odot$ & 0.25 & 0.17 \\ \hline
\end{tabular}
\caption{Same as \autoref{tab:sfh_area}, but for the total number of sinks/SSPs (\autoref{fig:num_stars1} and \autoref{fig:num_stars2}).}
\label{tab:nsinks_area}
\end{table}

When feedback is enabled, the \code{M2e7\_R300} starts forming SSPs later in the \ps run. However, the total number catches up quickly and eventually exceeds that of the run at natively high resolution. For the \code{M2e5\_R30} cloud with feedback, the \ps run ($\Delta m_i= 16 \rightarrow 1 \,\Msun$, orange dash-dotted line) produces roughly the same number of SSP particles as the run at natively high resolution until $\sim 1 \, \tff$. However, beyond this point the \ps run deviates and saturates at a value lower by an order of magnitude. This is because the \ps run forms both sinks and star particles and after $1\, \tff$ the $\approx 10$ massive sink particles disrupt the cloud as they go supernova (similar to the blue solid and purple dash-dotted line. The \ps run ($\Delta m_i= 128 \rightarrow 1 \,\Msun$, purple dash-dotted line) takes longer to split particles and has a smoother density structure owing to the lower starting resolution. As a result, it does not form any FIRE SSPs, which generally form more easily (given their lax formation criteria), when compared to sink particles. This run also forms roughly the same number of stars as the natively high resolution run (blue solid line). 

We summarize our results from \autoref{fig:num_stars1} and \autoref{fig:num_stars2} in \autoref{tab:nsinks_area}. In all cases, the ratios $A_{\Delta m}/A_{\rm 1\sigma}$ and $A_{\Delta m}/A_{\rm range}$ are less than unity. Once again we emphasize that in all cases, the differences we see in the runs with and without \ps are smaller compared to the differences in the runs without particle splitting where we vary the initial turbulent random seed. We therefore argue that particle splitting is not sampling the wrong solution space entirely.

\subsection{The IMF and Light-to-mass ratio}

\begin{figure}[!h]
    \centering
    \includegraphics[width=\linewidth]{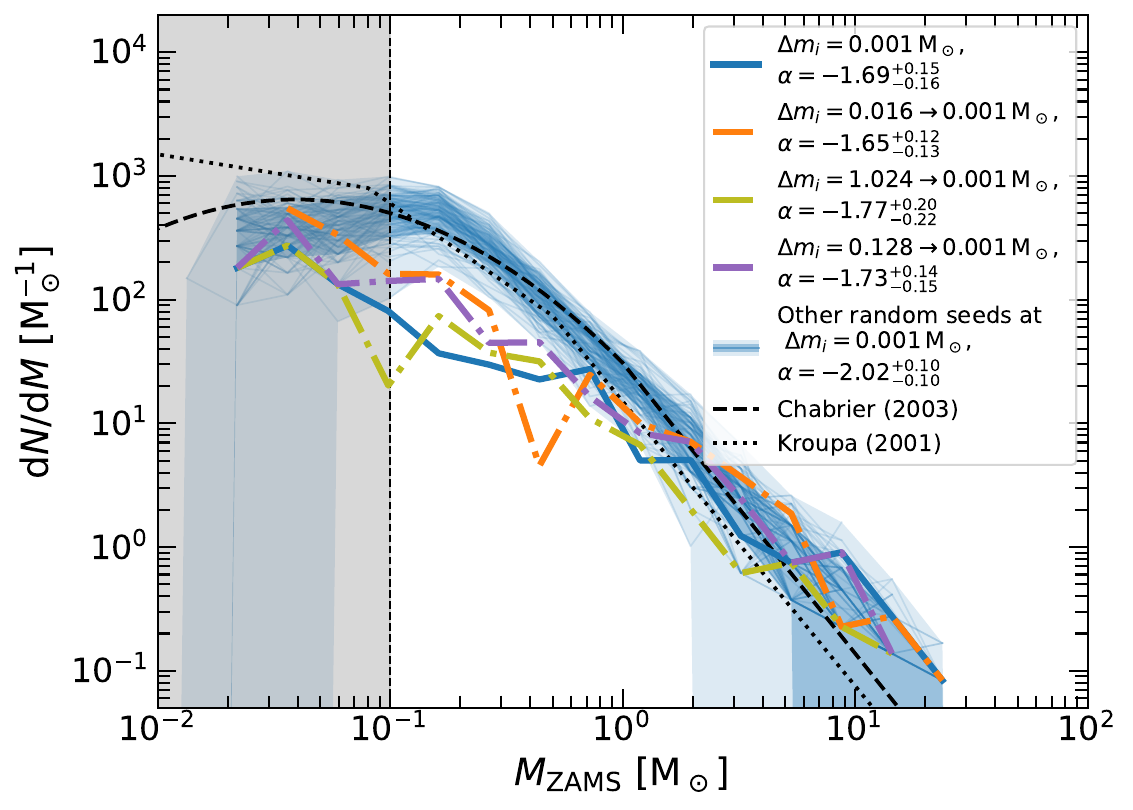}
    \caption{The IMF in the \code{M2e3\_R3} cloud runs with feedback. The variation between the mass spectrum of \ps runs (dash-dotted lines) and the natively high resolution run (thick blue solid lines) is roughly comparable to the scatter in the 99 runs (thin blue solid lines) where we varied the initial turbulent velocity field. The shaded bands denote the $16^{\rm th} - 84^{\rm th}$ percentile range (dark) and the entire range (light) for the random seed runs. The differences between the different random seed runs and the natively high resolution run are due to updates to the code version (see text for a detailed explanation).}
    \label{fig:imf}
\end{figure}

\begin{table}
\centering
\setlength{\tabcolsep}{2pt}
\begin{tabular}{cccc}
\toprule
Cloud & $\Delta m_i$ & $A_{\Delta m}/A_{\rm 1\sigma}$ & $A_{\Delta m}/A_{\rm range}$ \\ \hline
\multicolumn{4}{c}{\textbf{With feedback}} \\ \hline
\texttt{M2e3\_R3} & $1.024 \rightarrow 0.001 \, {\rm M}_\odot$ & 0.63 & 0.27 \\
 & $0.128 \rightarrow 0.001 \, {\rm M}_\odot$ & 0.52 & 0.23 \\
 & $0.016 \rightarrow 0.001 \, {\rm M}_\odot$ & 0.88 & 0.38 \\ \hline
\end{tabular}
\caption{Same as \autoref{tab:sfh_area}, but for the IMF in the \code{M2e3\_R3} cloud runs with feedback (\autoref{fig:imf}).}
\label{tab:imf_area}
\end{table}

In \autoref{sec:SFH} and \autoref{sec:num-stars}, we showed that the total mass and number of stars appears to be roughly similar when comparing \ps runs with runs at natively high resolution. However, this does not guarantee that the mass spectrum will be the same as well. Since cloud destruction is driven by feedback, even slight differences in the number of massive stars can potentially be important. In this section we investigate the IMF of stars in the simulations where we can resolve it ($\Delta m = 0.001 \, \Msun$).

\autoref{fig:imf} shows the IMF in \code{M2e3\_R3} clouds with feedback. The IMF in the \ps runs exhibits differences when compared to the run at natively high resolution, particularly between $0.1-0.5 \, \Msun$. We find that at the high mass end, these differences are smaller and in reasonable agreement with each other, with best fit values\footnote{\label{version-updates}We find the best-fit slope by 
fitting $dN/dM \propto M^{\alpha}$ within [$0.5\, \Msun, 200 \, \Msun$]. 
The fit is unbinned: we evaluate the likelihood for $\alpha \in [-4,4]$ and calculate the posterior $P(\alpha|\{m_i\})$ using a flat prior. We then quote the 16$^{\rm th}$, 50$^{\rm th}$ and 84$^{\rm th}$ percentiles of $P(\alpha|\{m_i\})$.} of $\alpha = -1.69^{+0.15}_{-0.16}$ for the natively high resolution run and $\alpha = -1.65^{+0.12}_{-0.13}$, $\alpha = -1.77^{+0.20}_{-0.22}$, and $\alpha = -1.73^{+0.14}_{-0.15}$ for the \ps runs.

When comparing to the set of simulations with 99 different random seeds (from \citet{Grudic2023b}, we find that the spread seen in these runs is larger than the differences seen between the \ps runs and the run at natively high resolution (see \autoref{tab:imf_area}). We note again that the 99 different random seed runs were run using a different code version. Compared to the earlier version, the IMF slope is more top-heavy because of updates to the thermodynamics, wind feedback physics, RT solver, as well-as several bug-fixes (see footnote \ref{version-updates}). We also note that the microphysics in the newer code version are improvements, despite the fact that the IMF deviates from the canonical \cite{Chabrier2003} or \citep{Kroupa2001} IMF.

\begin{figure}
    \centering
    \includegraphics[width=\linewidth]{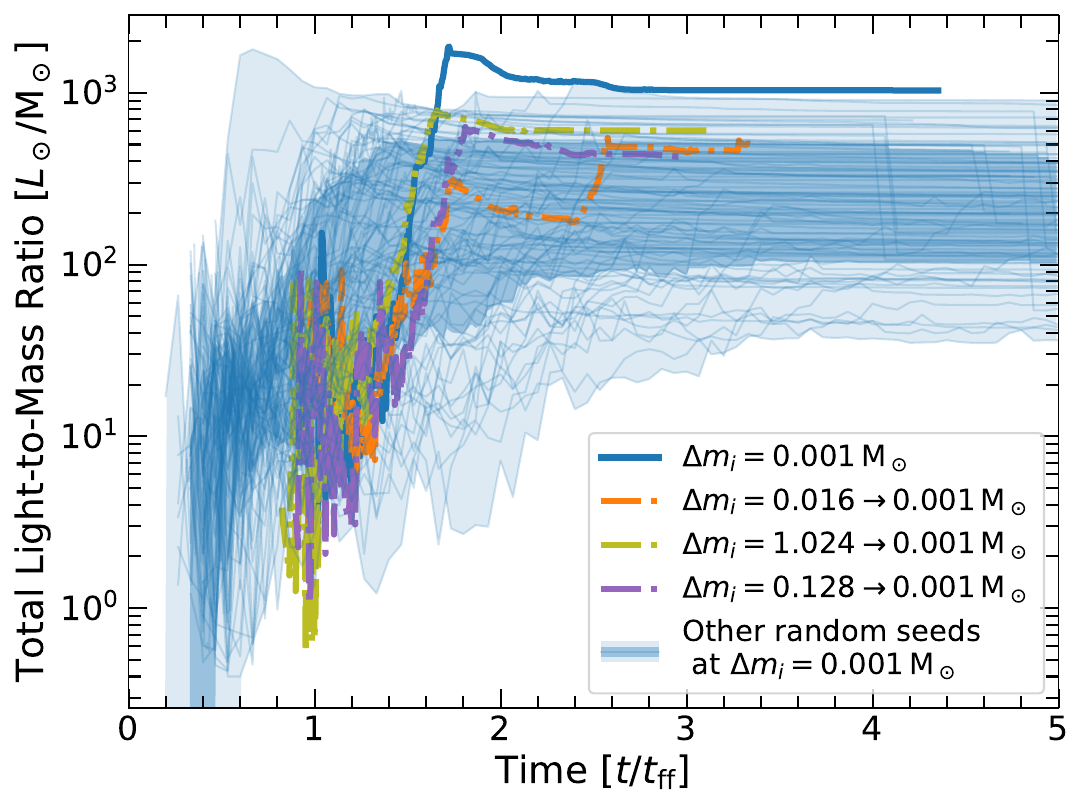}
    \caption{The total light-to-mass (L/M) ratio as a function of time in the \code{M2e3\_R3} cloud runs with feedback. Particle splitting runs are denoted by dash-dotted lines, the natively high resolution run is denoted by the thick blue solid line and the 99 different random seed runs are denoted by thin blue solid lines. The shaded bands denote the $16^{\rm th} - 84^{\rm th}$ percentile range (dark) and the entire range (light) for the random seed runs. The L/M ratio is variable at early times when the protostars are still forming and reaching the main sequence and \ps runs have a similar L/M ratio as the natively high resolution run. At late times, they diverge because of differences in the number of stars and their mass distribution. The differences between the \ps and natively high resolution run are within the range of values seen by varying the initial random seed.}
    \label{fig:light-to-mass}
\end{figure}

\begin{table}
\centering
\setlength{\tabcolsep}{2pt}
\begin{tabular}{cccc}
\toprule
Cloud & $\Delta m_i$ & $A_{\Delta m}/A_{\rm 1\sigma}$ & $A_{\Delta m}/A_{\rm range}$ \\ \hline
\multicolumn{4}{c}{\textbf{With feedback}} \\ \hline
\texttt{M2e3\_R3} & $1.024 \rightarrow 0.001 \, {\rm M}_\odot$ & 0.25 & 0.10 \\
 & $0.128 \rightarrow 0.001 \, {\rm M}_\odot$ & 0.29 & 0.12 \\
 & $0.016 \rightarrow 0.001 \, {\rm M}_\odot$ & 0.26 & 0.11 \\ \hline
\end{tabular}
\caption{Same as \autoref{tab:sfh_area}, but for the light-to-mass ratio in the \code{M2e3\_R3} cloud runs with feedback (\autoref{fig:light-to-mass}).}
\label{tab:lm_area}
\end{table}

\begin{figure*}
    \centering
    \includegraphics[width=\textwidth]{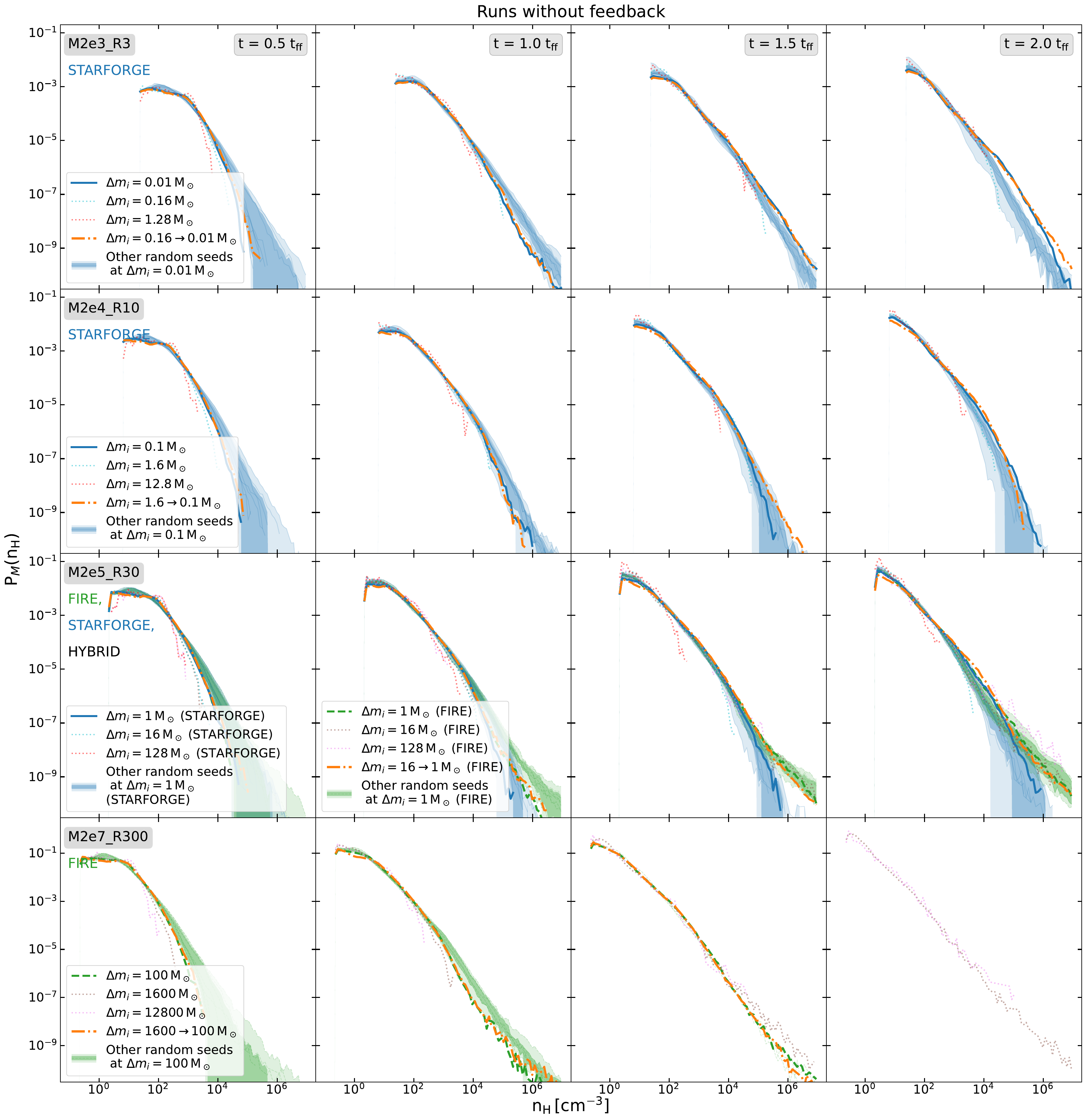}
    \caption{The density PDF, $P_M(\rm n_H)$, shown for 4 different times for runs without feedback. The density PDF shows a power-law tail at high densities in both runs with and without particle splitting. The differences between the \ps runs and those at natively resolution are negligible, especially when compared to the scatter present when changing the random seed.}
    \label{fig:pdf_nofb}
\end{figure*}

\begin{figure*}
    \centering
    \includegraphics[width=\textwidth]{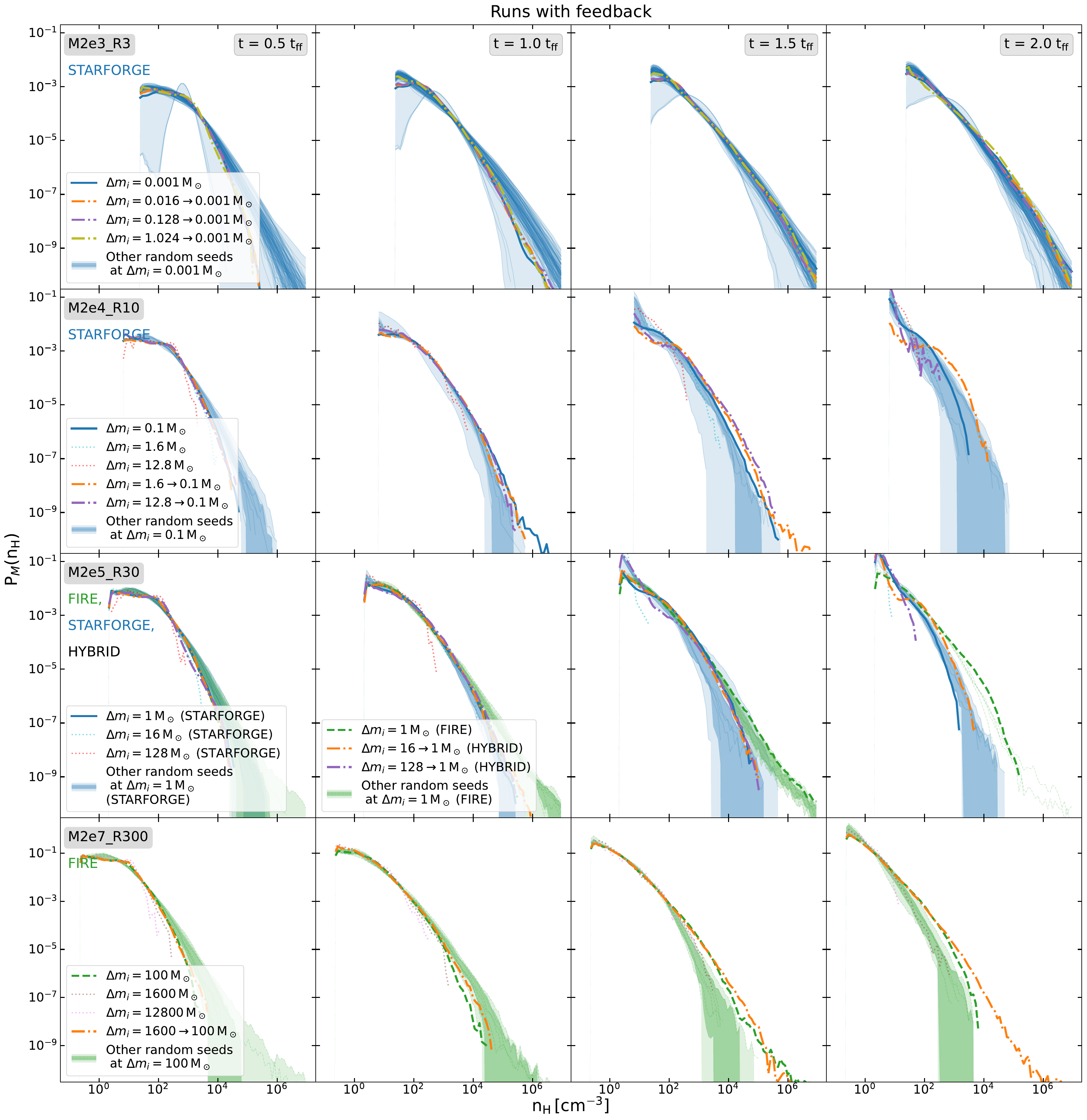}
    \caption[]{Same as \autoref{fig:pdf_nofb}, but for runs with feedback. At early times, \ps runs match the natively high resolution runs closely, but at late times, they can diverge. The differences between the \ps runs and those at uniformly \high resolution are smaller than or, at worst, comparable to the random seed scatter.}
    \label{fig:pdf_fb}
\end{figure*}

Cloud destruction in STARFORGE simulations begins with early-time radiation feedback \citep{Guszejnov2021, Grudic2022}. To quantify the impact of the differences in the IMF on the cloud destruction, we consider the ratio of the total bolometric luminosities of all the STARFORGE sink particles to their total mass in a given \code{M2e3\_R3} simulation ($L_{\star, \rm tot} $/$M_{\star, \rm tot}$) in \autoref{fig:light-to-mass}. We find that $L_{\star, \rm tot} $/$M_{\star, \rm tot}$ ranges over 2 orders of magnitude for the 99 different random seed runs (thin blue solid lines) at early times and over an order of magnitude at late times. For the run at natively high resolution (thick blue solid line), this ratio is higher once stars have stopped forming ($\sim 2 \, \tff$. This is not surprising given that the IMF in this run is comparably more top-heavy. For the \ps runs $\Delta m_i= 0.128 \rightarrow 0.001 \,\Msun$ (purple dash-dotted line) and $\Delta m_i= 0.016 \rightarrow 0.001 \,\Msun$ (orange dash-dotted line), the ratio is lower and is explained by the presence of more low-mass stars in those runs. For the \ps run $\Delta m_i= 1.024 \rightarrow 0.001 \,\Msun$ (olive dash-dotted line), the ratio is lowered because of the lack of high mass stars when compared to the natively high resolution run.

We find that the differences between the \ps runs and the run at natively high resolution are smaller than the inherent scatter in this quantity because of the choice of the initial conditions as seen in \autoref{tab:lm_area}. We therefore argue that these light-to-mass ratio variations, along with the IMF and SFH variations, imply that \ps does not change the star formation outcome drastically beyond the ``acceptable" bounds in place due to the chaotic nature of the system and the choice of the initial conditions.

\subsection{The Density PDF and power spectrum}

Having analyzed the star formation outcome, we now turn to the density statistics of our clouds. In this section, we examine the density PDF and density power spectrum for all our runs to further understand the effects of splitting particles on the density structure of the gas. 

Our clouds are initialized at a constant density $\rho_0$, and embedded in a warm diffuse medium ($\rho_0/1000$) that fills the box out to $10 \,R_{\rm cloud}$. To generate the density PDF, we apply a density cut, selecting gas above the geometric mean of these two densities ($\rho>\rho_0/\sqrt{1000}$), and plot the mass-weighted number density PDF, $P_M(\rm n_H)$.

In \autoref{fig:pdf_nofb}, we plot the density PDF for the runs without feedback. For the \code{M2e3\_R3} cloud (top row), we find that the density PDFs in the natively high resolution run and the \ps run follow each other closely at all times. For the \code{M2e4\_R10} cloud, the runs with and without splitting diverge at high densities after $\sim 1 \, \tff$. This is likely due to differences in the number of sink particles formed  in the two runs, which then influence the high density tail of the density PDF.

For all the other clouds, the density PDF in the \ps runs also follow the PDF in the corresponding natively high resolution run closely at early times. At late times, they diverge slightly, since the exact sink/SSP mass distribution can alter the density PDF \citep{Murray2015, Murray2017, Khullar2021}. However, the differences between the two are either comparable to, or smaller than, the variation among different random seed runs (\autoref{tab:denspdf_area}).

\autoref{fig:pdf_fb} shows the density PDF in the runs with feedback. We find that in this case, for the \code{M2e3\_R3} clouds which have the largest number of particles in our simulation set, the density PDF is virtually identical in all our runs, regardless of whether or not we split particles. For the remaining cloud cases, we also find good agreement between splitting and non-splitting runs, regardless of the star particle formation recipe used. Once again, the scatter seen among the different random seed runs is larger than the differences seen between the runs where we split particles and runs where we have natively high resolution, as noted in \autoref{tab:denspdf_area}.

The signatures seen in the density PDF are also seen in the density power spectra. To be consistent across the varying sizes of clouds, we show the integer wavenumber $k_{\rm int}$ on the x-axis. To generate the power spectra, we choose a box size of $2\,R_{\rm cloud}$ centered at the cloud's center and interpolate particle data on to a $256^3$ grid. We caution the reader that the minimum distance between two particles is typically much smaller than our sampling grid resolution and matching the two is currently beyond computational capabilities. Because of this, the scales at which \ps and non \ps runs would be expected to deviate, are averaged over.

\begin{table}
\centering
\setlength{\tabcolsep}{2pt}
\begin{tabular}{ccccc}
\toprule
Cloud & $\Delta m_i$ & Time [$t_{\rm ff}$] & $A_{\Delta m}/A_{\rm 1\sigma}$ & $A_{\Delta m}/A_{\rm range}$ \\ \hline
\multicolumn{5}{c}{\textbf{With feedback}} \\ \hline
\texttt{M2e3\_R3} & $1.024 \rightarrow 0.001 \, {\rm M}_\odot$ & 0.5 & 0.29 & 0.05 \\
 &  & 1 & 0.25 & 0.07 \\
 &  & 1.5 & 0.27 & 0.08 \\
 &  & 2 & 0.41 & 0.13 \\
 & $0.128 \rightarrow 0.001 \, {\rm M}_\odot$ & 0.5 & 0.18 & 0.03 \\
 &  & 1 & 0.16 & 0.05 \\
 &  & 1.5 & 0.32 & 0.09 \\
 &  & 2 & 0.17 & 0.05 \\
 & $0.016 \rightarrow 0.001 \, {\rm M}_\odot$ & 0.5 & 0.09 & 0.02 \\
 &  & 1 & 0.13 & 0.04 \\
 &  & 1.5 & 0.27 & 0.08 \\
 &  & 2 & 0.32 & 0.10 \\
\texttt{M2e4\_R10} & $12.8 \rightarrow 0.1 \, {\rm M}_\odot$ & 0.5 & 0.19 & 0.12 \\
 &  & 1 & 0.52 & 0.21 \\
 &  & 1.5 & 0.47 & 0.18 \\
 &  & 2 & 0.71 & 0.50 \\
 & $1.6 \rightarrow 0.1 \, {\rm M}_\odot$ & 0.5 & 0.23 & 0.14 \\
 &  & 1 & 0.37 & 0.15 \\
 &  & 1.5 & 0.45 & 0.17 \\
 &  & 2 & 0.51 & 0.36 \\
\texttt{M2e5\_R30} & \multicolumn{4}{c}{\ldots} \\
\texttt{M2e7\_R300} & $1600 \rightarrow 100 \, {\rm M}_\odot$ & 0.5 & 0.22 & 0.14 \\
 &  & 1 & 0.35 & 0.25 \\
 &  & 1.5 & 0.28 & 0.17 \\
 &  & 2 & 0.98 & 0.61 \\ \hline
\multicolumn{5}{c}{\textbf{Without feedback}} \\ \hline
\texttt{M2e3\_R3} & $0.16 \rightarrow 0.01 \, {\rm M}_\odot$ & 0.5 & 0.29 & 0.21 \\
 &  & 1 & 0.21 & 0.12 \\
 &  & 1.5 & 0.16 & 0.09 \\
 &  & 2 & 0.38 & 0.19 \\
\texttt{M2e4\_R10} & $1.6 \rightarrow 0.1 \, {\rm M}_\odot$ & 0.5 & 0.15 & 0.10 \\
 &  & 1 & 0.33 & 0.16 \\
 &  & 1.5 & 0.63 & 0.38 \\
 &  & 2 & 0.41 & 0.22 \\
\texttt{M2e5\_R30} & $16 \rightarrow 1 \, {\rm M}_\odot$ & 0.5 & 0.31 & 0.20 \\
 &  & 1 & 0.26 & 0.14 \\
 &  & 1.5 & 0.35 & 0.21 \\
 &  & 2 & 0.28 & 0.17 \\
\texttt{M2e7\_R300} & $1600 \rightarrow 100 \, {\rm M}_\odot$ & 0.5 & 0.16 & 0.11 \\
 &  & 1 & 0.46 & 0.24 \\ \hline
\end{tabular}
\caption{Summary statistics for deviations in the mass-weighted density PDF, $P_M(n_{\rm H})$, while splitting particles (\autoref{fig:dens_ps_nofb} and \autoref{fig:dens_ps_fb}). Columns (1-3): Cloud case, resolution and time (in units of free-fall time). Column (4-5): The ratio of area between the run with and without particle splitting ($A_{\Delta m}$; \autoref{eq:A-dm}) to the seed-to-seed scatter area, measured against the $16^{\rm th}-84^{\rm th}$ percentile band ($A_{\rm 1\sigma}$; \autoref{eq:A-sigma}) and the full range ($A_{\rm range}$). All the ratios are again less than unity, indicating that the differences caused by particle splitting are less than the random-seed scatter.}
\label{tab:denspdf_area}
\end{table}

\begin{figure*}
    \centering
    \includegraphics[width=\textwidth]{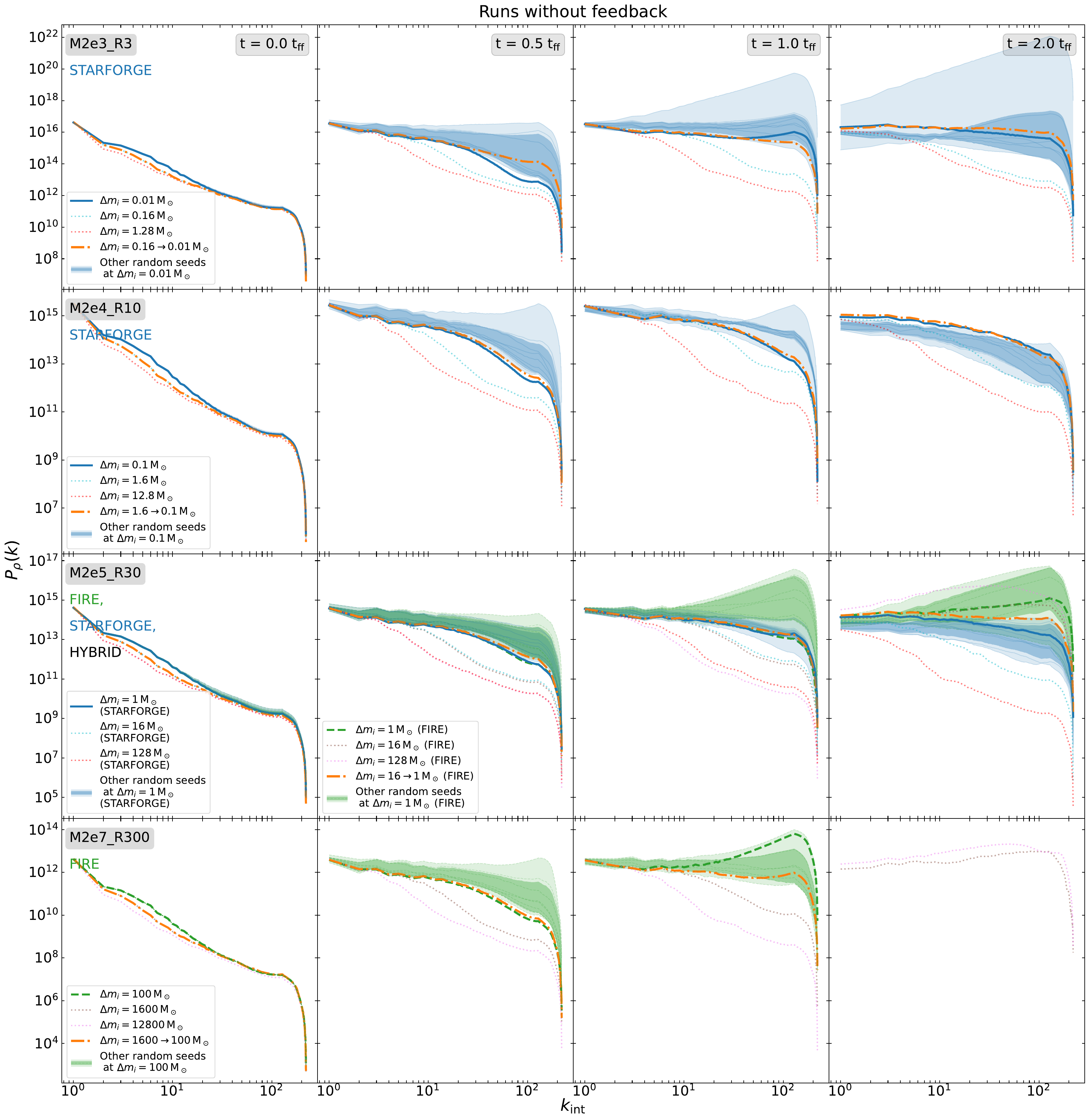}
    \caption{The density power spectrum, $P_\rho(k)$ as a function of the integer wavenumber $k_{\rm int}$ for runs without feedback. Particle splitting runs generally start off with lower power at large scales owing to the lower resolution. However, shortly afterwards, they match the power spectrum in the natively high resolution runs such that the differences between the two are smaller than the random seed scatter.}
    \label{fig:dens_ps_nofb}
\end{figure*}

\begin{figure*}
    \centering
    \includegraphics[width=\textwidth]{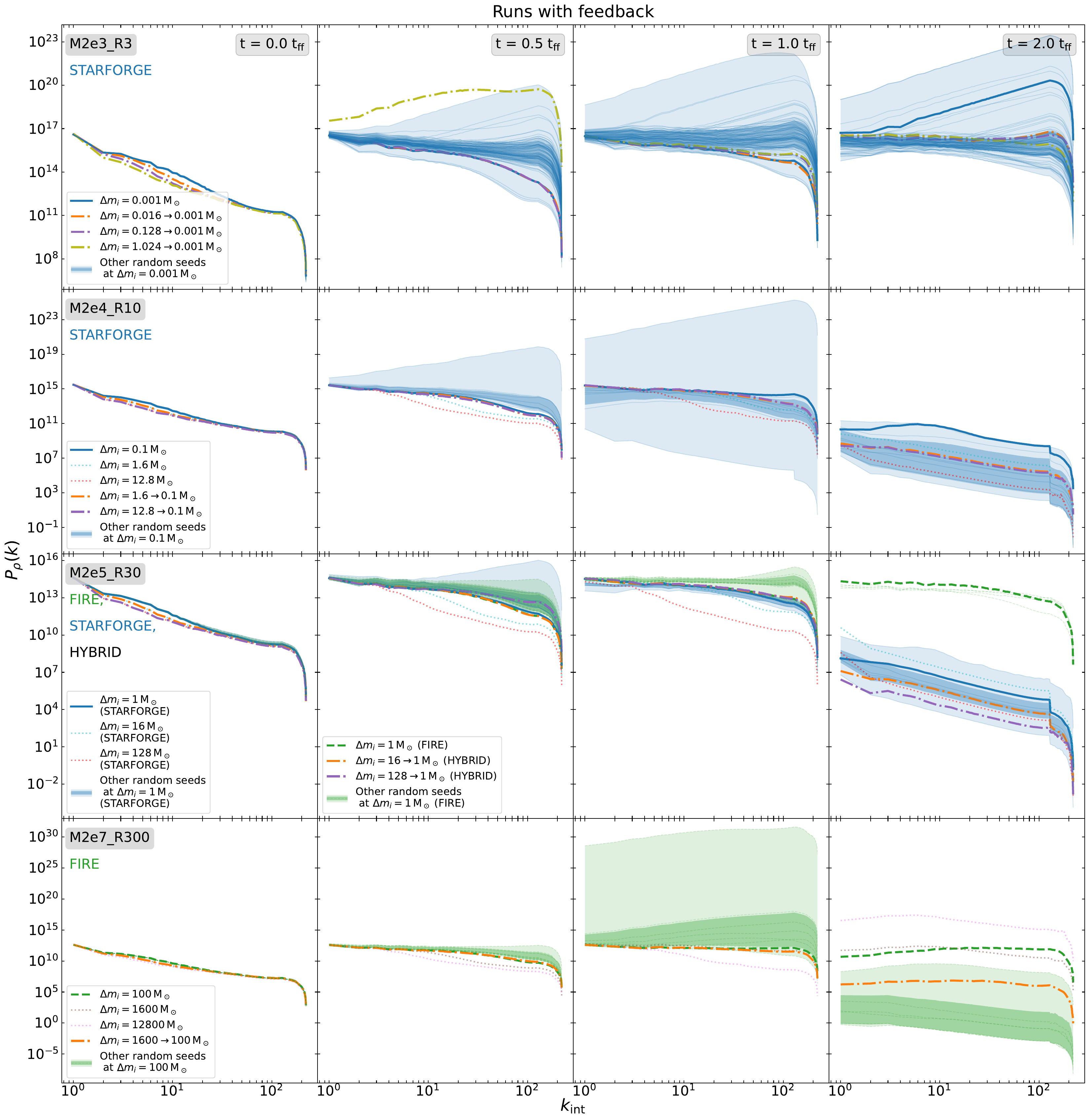}
    \caption{Same as \autoref{fig:dens_ps_nofb}, but for runs with feedback. The differences between the \ps and the natively high resolution runs are negligible at early times. The density power spectra for the \ps runs diverge from the natively high resolution run once feedback kicks in. The variation between the splitting and non-splitting runs is still smaller than the range seen for different random seed runs.}
    \label{fig:dens_ps_fb}
\end{figure*}

We first consider the density power spectrum in our runs without feedback in \autoref{fig:dens_ps_nofb}. At $t=0$, none of the \ps runs match the natively high resolution run since they start off at a lower resolution. For the \code{M2e3\_R3} cloud (top row), the power spectra are similar in the \ps and natively high resolution runs once the \ps run has reached the same resolution. For the \code{M2e4\_R10} case, the differences in the power spectra of the \ps and the natively high resolution runs are similarly negligible. However, for the \code{M2e5\_R30} and \code{M2e7\_R300} clouds, the power spectra diverge by $2 \, \tff$ and $1 \, \tff$ respectively. Nonetheless, in all cloud cases and at all times, the differences between the \ps and natively high resolution run are smaller than the differences seen between the runs with different random seeds ($A_{\Delta m}/A_{\rm range}<1$).

In \autoref{fig:dens_ps_fb}, we show the density power spectra for the runs with feedback. By $0.5 \, \tff$, the \code{M2e3\_R3} clouds show a wide spread for the different random seed runs (thin blue solid lines). The power spectrum in the \ps runs are identical to the natively high resolution run, except for the $\Delta m_i= 1.024 \rightarrow 0.001 \, \Msun$ run (olive dash-dotted line). This deviation is transient as the $\Delta m_i= 1.024 \rightarrow 0.001 \, \Msun$ run drops down close to the natively high resolution run by $1 \, \tff$. 

 \begin{table}
\centering
\setlength{\tabcolsep}{2pt}
\begin{tabular}{ccccc}
\toprule
Cloud & $\Delta m_i$ & Time [$t_{\rm ff}$] & $A_{\Delta m}/A_{\rm 1\sigma}$ & $A_{\Delta m}/A_{\rm range}$ \\ \hline
\multicolumn{5}{c}{\textbf{With feedback}} \\ \hline
\texttt{M2e3\_R3} & $1.024 \rightarrow 0.001 \, {\rm M}_\odot$ & 0.5 & 5.28 & 0.94 \\
 &  & 1 & 0.21 & 0.04 \\
 &  & 2 & 2.37 & 0.32 \\
 & $0.128 \rightarrow 0.001 \, {\rm M}_\odot$ & 0.5 & 0.04 & 0.01 \\
 &  & 1 & 0.14 & 0.03 \\
 &  & 2 & 2.27 & 0.31 \\
 & $0.016 \rightarrow 0.001 \, {\rm M}_\odot$ & 0.5 & 0.03 & 0.01 \\
 &  & 1 & 0.05 & 0.01 \\
 &  & 2 & 2.16 & 0.30 \\
\texttt{M2e4\_R10} & $12.8 \rightarrow 0.1 \, {\rm M}_\odot$ & 0.5 & 0.17 & 0.03 \\
 &  & 1 & 0.19 & 0.02 \\
 &  & 2 & 0.76 & 0.46 \\
 & $1.6 \rightarrow 0.1 \, {\rm M}_\odot$ & 0.5 & 0.10 & 0.02 \\
 &  & 1 & 0.21 & 0.02 \\
 &  & 2 & 0.73 & 0.44 \\
\texttt{M2e5\_R30} & \multicolumn{4}{c}{\ldots} \\
\texttt{M2e7\_R300} & $1600 \rightarrow 100 \, {\rm M}_\odot$ & 0.5 & 0.09 & 0.05 \\
 &  & 1 & 0.05 & 0.01 \\
 &  & 2 & 1.11 & 0.50 \\ \hline
\multicolumn{5}{c}{\textbf{Without feedback}} \\ \hline
\texttt{M2e3\_R3} & $0.16 \rightarrow 0.01 \, {\rm M}_\odot$ & 0.5 & 0.23 & 0.16 \\
 &  & 1 & 0.34 & 0.12 \\
 &  & 2 & 0.90 & 0.23 \\
\texttt{M2e4\_R10} & $1.6 \rightarrow 0.1 \, {\rm M}_\odot$ & 0.5 & 0.14 & 0.07 \\
 &  & 1 & 0.23 & 0.06 \\
 &  & 2 & 0.11 & 0.07 \\
\texttt{M2e5\_R30} & $16 \rightarrow 1 \, {\rm M}_\odot$ & 0.5 & 0.12 & 0.08 \\
 &  & 1 & 0.09 & 0.06 \\
 &  & 2 & 0.52 & 0.27 \\
\texttt{M2e7\_R300} & $1600 \rightarrow 100 \, {\rm M}_\odot$ & 0.5 & 0.09 & 0.05 \\
 &  & 1 & 0.93 & 0.56 \\ \hline
\end{tabular}
\caption{Same as \autoref{tab:denspdf_area}, but for the density power spectrum $P_\rho(k)$.}
\label{tab:densps_area}
\end{table}

For the \code{M2e4\_R10} cloud, the \ps runs are close to the natively high resolution until $1 \, \tff$. By $2 \, \tff$, the density power spectra diverge as the cloud is destroyed. A similar trend is seen for the \code{M2e5\_R30} and \code{M2e7\_R300} clouds as well. The \code{M2e5\_R30} shows a large difference in the density power spectrum when we use the FIRE star formation prescription versus when we use the STARFORGE prescription. This is due to the more energetic feedback caused by the massive sinks (as a result of the gas mass resolution of $\Delta m = 1\, \Msun$) in the STARFORGE run (blue solid line). The \ps runs (orange and purple dash-dotted lines) also undergo destruction by massive sinks. The \ps runs have a mix of FIRE SSPs and STARFORGE sinks since FIRE SSPs form when particles are still being split. 

We summarize these results in \autoref{tab:densps_area}. At late times, $A_{\Delta m}/A_{1\sigma}$ increases and even exceeds unity for some cases. This is caused by feedback altering the density distribution non-linearly and destroying the cloud. However, even for these cases, $A_{\Delta m}/A_{\rm range}<1$, so the \ps runs do not deviate beyond the plausible range. 

\subsection{Velocity power spectrum}

\begin{figure*}
    \centering
    \includegraphics[width=\textwidth]{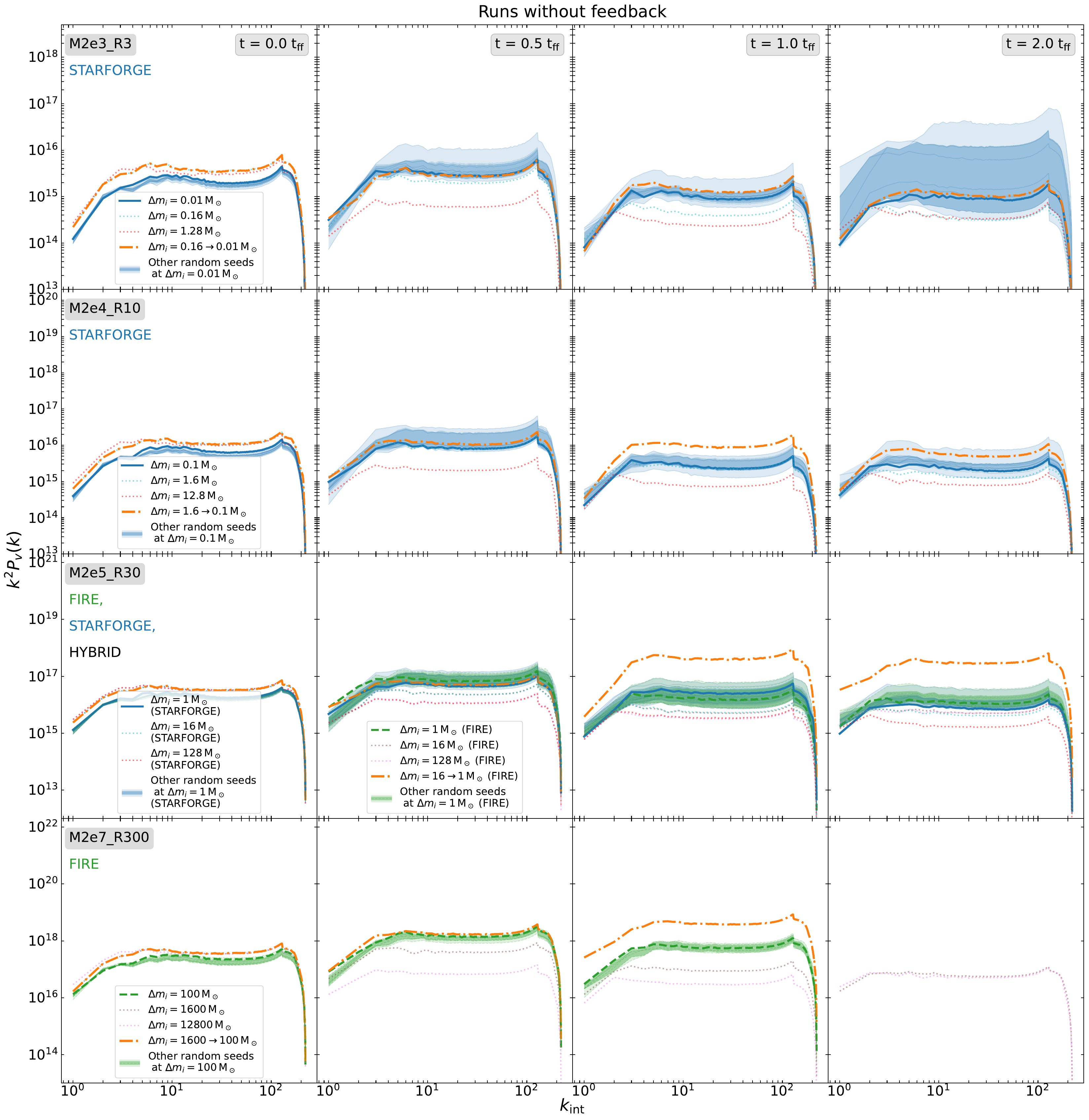}
    \caption{The velocity power spectrum, normalized by $k^{2}$ as a function of the integer wavenumber $k_{\rm int}$ for runs without feedback. 
    Runs with \ps start out with slightly higher power than the natively high resolution runs but match them shortly thereafter. They deviate in magnitude after $1\tff$, but the differences between the \ps and the natively resolution runs are comparable to the random seed scatter.}
    \label{fig:vel_ps_nofb}
\end{figure*}

\begin{figure*}
    \centering
    \includegraphics[width=\textwidth]{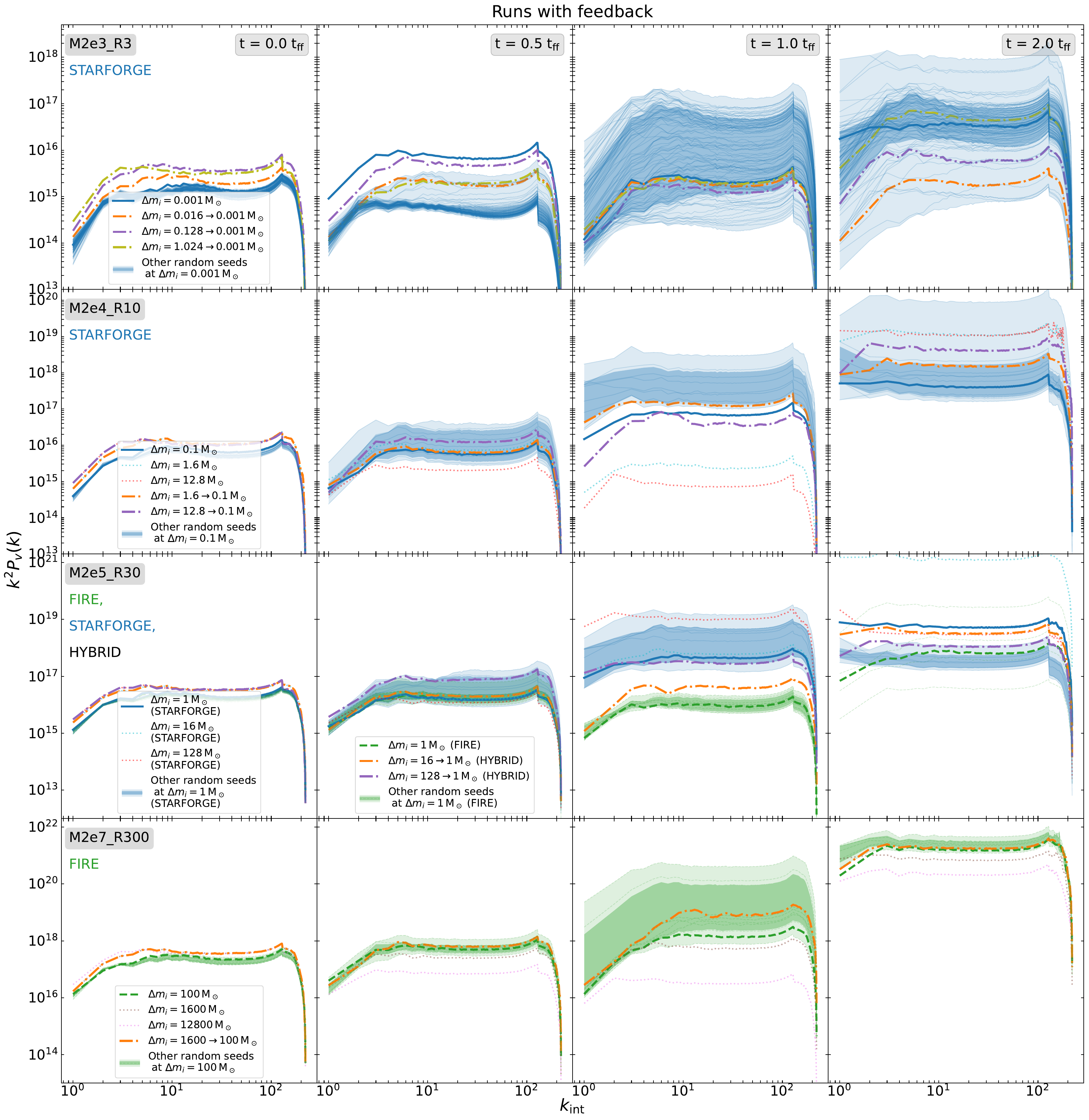}
    \caption{Same as \autoref{fig:vel_ps_nofb}, but for runs with feedback. Feedback eventually leads to a large spread in the magnitude of the power spectrum at late times, but the differences between the \ps and the natively resolution runs are again comparable to or smaller than the random seed scatter.}
    \label{fig:vel_ps_fb}
\end{figure*}

\begin{table}
\centering
\setlength{\tabcolsep}{2pt}
\begin{tabular}{ccccc}
\toprule
Cloud & $\Delta m_i$ & Time [$t_{\rm ff}$] & $A_{\Delta m}/A_{\rm 1\sigma}$ & $A_{\Delta m}/A_{\rm range}$ \\ \hline
\multicolumn{5}{c}{\textbf{With feedback}} \\ \hline
\texttt{M2e3\_R3} & $1.024 \rightarrow 0.001 \, {\rm M}_\odot$ & 0.5 & 3.74 & 0.79 \\
 &  & 1 & 0.03 & 0.02 \\
 &  & 2 & 0.23 & 0.06 \\
 & $0.128 \rightarrow 0.001 \, {\rm M}_\odot$ & 0.5 & 1.31 & 0.28 \\
 &  & 1 & 0.12 & 0.08 \\
 &  & 2 & 0.85 & 0.22 \\
 & $0.016 \rightarrow 0.001 \, {\rm M}_\odot$ & 0.5 & 3.58 & 0.76 \\
 &  & 1 & 0.04 & 0.02 \\
 &  & 2 & 1.53 & 0.40 \\
\texttt{M2e4\_R10} & $12.8 \rightarrow 0.1 \, {\rm M}_\odot$ & 0.5 & 0.42 & 0.28 \\
 &  & 1 & 0.28 & 0.17 \\
 &  & 2 & 1.13 & 0.36 \\
 & $1.6 \rightarrow 0.1 \, {\rm M}_\odot$ & 0.5 & 0.09 & 0.06 \\
 &  & 1 & 0.33 & 0.20 \\
 &  & 2 & 0.62 & 0.20 \\
\texttt{M2e5\_R30} & \multicolumn{4}{c}{\ldots} \\
\texttt{M2e7\_R300} & $1600 \rightarrow 100 \, {\rm M}_\odot$ & 0.5 & 0.41 & 0.19 \\
 &  & 1 & 0.34 & 0.21 \\
 &  & 2 & 0.27 & 0.15 \\ \hline
\multicolumn{5}{c}{\textbf{Without feedback}} \\ \hline
\texttt{M2e3\_R3} & $0.16 \rightarrow 0.01 \, {\rm M}_\odot$ & 0.5 & 0.59 & 0.26 \\
 &  & 1 & 2.10 & 0.80 \\
 &  & 2 & 0.04 & 0.03 \\
\texttt{M2e4\_R10} & $1.6 \rightarrow 0.1 \, {\rm M}_\odot$ & 0.5 & 0.30 & 0.20 \\
 &  & 1 & 1.85 & 0.98 \\
 &  & 2 & 1.62 & 0.47 \\
\texttt{M2e5\_R30} & $16 \rightarrow 1 \, {\rm M}_\odot$ & 0.5 & 0.38 & 0.26 \\
 &  & 1 & 2.43 & 1.57 \\
 &  & 2 & 3.41 & 1.84 \\
\texttt{M2e7\_R300} & $1600 \rightarrow 100 \, {\rm M}_\odot$ & 0.5 & 0.50 & 0.28 \\
 &  & 1 & 3.07 & 2.27 \\ \hline
\end{tabular}
\caption{Same as \autoref{tab:denspdf_area}, but for the velocity power spectrum, $k^2 P_v(k)$.}
\label{tab:velps_area}
\end{table}

In this section, we consider the velocity power spectrum. We calculate the velocity power spectrum in the same manner as the density power spectrum. We further normalize the velocity power spectrum by the $k^{2}$ to ensure that the $k^{-2}$ Burgers-like scaling seen in supersonic turbulence appears parallel to the x-axis. 

In the runs without feedback (\autoref{fig:vel_ps_nofb}), the \ps runs have more power at $t=0$ when compared to the natively high resolution runs (for all cloud cases). This is due to the different initial resolutions: runs with the same initial resolution and random seeds indeed have identical power spectra. 
However, by $t=0.5 \, \tff$ the velocity power spectra curves for the \ps and natively high resolution runs lie very close to each other, with the difference being within the scatter present in the different random seeds at high resolution. The curves diverge by $t=1 \, \tff$, and \ps runs tend to have a systematically higher velocity dispersion than the natively high resolution runs. We also note that the magnitude of the velocities do not change significantly over time in runs without feedback.

In \autoref{fig:vel_ps_fb}, we show the velocity power spectra for runs with feedback. For the \code{M2e3\_R3} cloud, the \ps runs are close to the run with natively high resolution until $1 \, \tff$. After this time, the curves diverge and the magnitude of the velocities generally increase owing to the effects of feedback and cloud destruction. The same trend is observed in the remaining cloud cases as well, albeit in some cases, the \ps runs' power spectra remains close to that of the natively high resolution run even at late times. 
We note again that the differences caused by \ps are typically less than the differences caused by the choice of the random seed for the initial conditions. However, we find that, in general, \ps runs tend to have more power on all scales when compared to their natively high resolution counterparts. 

We report the $A_{\Delta m}$, $A_{1\sigma}$ and $A_{\rm range}$ values as ratios in \autoref{tab:velps_area}. Our ``weak convergence" criteria still holds in general, but is indeed violated for 3 cases, \code{M2e5\_R30} at $1\, \tff$ and $2\, \tff$ and \code{M2e7\_R300} at $1\, \tff$. We speculate that this may be stochastic in nature, since the condition is only violated at late times, once star formation is well underway.

\newpage
\section{Summary and Conclusions}
\label{sec:Summary}

In this paper, we perform $\approx 220$ simulations varying the cloud masses, radii, star particle formation algorithm, feedback physics and details of the splitting algorithm. We study the star formation outcomes of these simulations to understand the effects of \ps on fragmentation. We examine the density PDF and the density and velocity power spectra to understand the effects of \ps on the quantities associated with the turbulence in GMCs. We also perform a set of additional tests to check for artificial fragmentation in \autoref{sec:truelove}.

Our main conclusions can be summarized as follows:

\begin{itemize}
    \item \textit{Does particle splitting seed artificial fragmentation?} Our tests of an isothermal rotating collapsing cloud in \autoref{sec:truelove} show that \ps does not introduce artificial fragmentation. 
    For the turbulent sphere runs (main text), we can form a different number of STARFORGE sink or FIRE SSP particles. This is due to the non-linear chaotic nature of the system, where small perturbations can be amplified and lead to diverging solutions. Nevertheless, we argue that the differences introduced by \ps are within the scatter inherently present in these systems when the random seed used to generate the initial velocity field is varied.

    \item \textit{Does particle splitting alter the star formation outcome?} While the number of sinks or stars can be different in \ps runs (primarily when the STARFORGE sink formation algorithm is used), they nevertheless end up producing a similar total stellar mass compared to the runs where we maintain natively high resolution. 
    While particle splitting runs do not reproduce the exact star formation outcome when compared to the high resolution runs without splitting, they nevertheless produce a SFH that is within the range of SFHs of other random seed runs. Since the same stellar mass is produced by a different total number of stars, the IMF and light-to-mass ratio is also affected when we split particles. However, once again the differences in the IMF and light-to-mass ratio are within the scatter inherently present due to the choice of the initial conditions. Thus we conclude that \ps does not change the star formation outcome significantly.

    \item \textit{Do particle splitting runs generate the right turbulence statistics?} 
    The differences in the density PDF between the \ps runs and the corresponding runs at natively high resolution are within the scatter that exists between different random seed runs. The same is true for the density power spectra, with negligible differences caused by splitting particles. The late-time evolution of the density power spectra can different when feedback is enabled. This is due to effects of feedback, destroying the cloud at slightly different times because of differences in the masses of the sinks when we split particles.

    The velocity power spectra follow a $k^{-2}$ scaling and \ps runs do not show meaningful differences until later in the clouds' evolution. However, particle splitting runs in general have more power on all scales when compared to their corresponding natively high resolution runs.
    In runs with feedback, the scatter in the velocity spectra across different random seeds (once stars start disrupting the cloud) is greater than the differences between \ps and natively resolution runs.

\end{itemize}

We argue that overall, while a particle splitting approach introduces numerical errors when compared to the natively high resolution simulation, these errors are not generally larger than the variance due to the arbitrary choice of the initial random seed. For certain quantities of interest, e.g. the star formation history and the IMF, particle splitting simulations do not necessarily reduce the precision significantly since the uncertainties are dominated by stochasticity anyway.

Our results here imply that \ps generally works for star formation applications. However, it is not perfect and there might be ways to improve the method and converge to the high resolution run more accurately for fluid flows in various kinds of astrophysical systems. We also caution that neither the FIRE SSP particle approach nor the STARFORGE sink particle approach is necessarily the correct approach for a star/sink formation algorithm in the regime where the gas mass resolution is $\Delta m_i = 0.1-10\, \Msun$. This is especially important for studies where \ps is used to increase resolution and refine from large kpc scales down to sub-pc or sub-AU scales \citep[e.g.][]{Meziani2026}. We leave these improvements to future work.

\section*{Acknowledgements}

We thank Mordecai-Mark Mac-Low for comments on the manuscript. We also thank James Beattie, Yanlong Shi, Lachlan Lancaster, Stella Offner, Marta Reina-Campos, Raghav Arora and Shyam Menon for useful discussions.  The Flatiron Institute is a division of the Simons foundation. The research of SK, CM, and NM was supported by Discovery Grants from the National Science and Engineering Research Council.  NM was supported by a Canada Research Chair (Tier 1). Computations were performed partly on the Niagara/Trillium supercomputer at the SciNet HPC Consortium. SciNet is funded by Innovation, Science and Economic Development Canada; the Digital Research Alliance of Canada; the Ontario Research Fund: Research Excellence; and the University of Toronto. P.F.H. acknowledges support from a Simons Investigator
Grant. CAFG was supported by NSF through grants AST-2108230, AST-2307327 and AST-2606015; by NASA through grants 80NSSC22k0809, 80NSSC22K1124 and 80NSSC24K1224; by STScI through grant JWST-AR-03252.001-A; and by BSF through grant number 2024262.


\appendix

\section{Jeans-based refinement criterion tests}
\label{sec:truelove}

In a seminal paper, \citet{Truelove1997} show that perturbations arising from the grid discretization can grow and cause artificial fragmentation if the Jeans condition is not satisfied by at least 4 grid cells. Artificial fragmentation occurs when there is a mismatch between the (required) gravitational softening and the hydrodynamics (pressure) resolution, and unstable gravitational modes grow faster than the pressure forces can equalize. SPH-based methods do not suffer from artificial fragmentation \citep{Bate1997, Whitworth1998, Hubber2006, Grudic2021} even when the Jeans length is poorly resolved. This is partly because in these methods, the gravitational softening is often tied to the smoothing length. Hence, when gas gets denser, particles crowd together and both the gravitational softening and smoothing length decrease, and the mismatch does not occur. 
However, it is not clear if the density perturbations introduced by hyper-Lagrangian refinement cause artificial fragmentation to occur. We therefore re-visit and validate our particle splitting method against established tests of artificial fragmentation in this section.

Similar to \cite{Truelove1997}, we perform the \cite{Boss1979} test using the same initial conditions: a $1\, \Msun$ cloud with a radius ${\rm R} = 0.0162 \, {\rm pc}$, and a Gaussian density profile. We apply an initial $m=2$ perturbation as: $\rho(r) \rightarrow \rho(r) (1+0.1 \cos 2\phi)$, where $\phi$ is the azimuthal angle, and follow the subsequent collapse until gas reaches a maximum density of $\rho_{\rm max} = 10^{-9.5}\, {\rm g \, cm^{-3}}$. To compare with grid-based methods, we define our equivalent cell-size as $\Delta x = (\Delta m/\rho)^{1/3}$. The Jeans length is calculated as $\lambda_J (\rho) = \sqrt{\pi c_s^2/G\rho}$. We define three different Jeans numbers to characterize our runs.
\begin{enumerate}
    \item The initial Jeans number: 
    \begin{equation}
        J_{\rm init} = \frac{\Delta x_{\rm init}}{\lambda_{J, \rm init}},
    \end{equation}
    where $\Delta x_{\rm init}$ is the initial grid cell-size, $\lambda_{J, \rm init}$ is the initial Jeans length using the mean density of the cloud.
    \item The refinement Jeans number:
    \begin{equation}
    \label{eq:refinement-jeans-number}
        J = J_{\rm ref} = \frac{\Delta x}{\lambda_{J}},
    \end{equation}
    where $\Delta x$ is the cell-size at any given refinement level and $\lambda_J\equiv\lambda_J(\rho)$. We drop the subscript below and refer to this as the Jeans number, $J$, since this is the commonly used definition in the literature. 
    \item The maximum Jeans number:
    \begin{equation}
        J_{\rm max} = \frac{\Delta x_{\rm min}}{\lambda_{J, \rm min}}.
    \end{equation}
    where $\Delta x_{\rm min}$ is the minimum cell-size and $\lambda_{J, \rm min}$ is the minimum Jeans length achieved at the maximum density $\rho_{\rm max} = 10^{-9.5}\, {\rm g \, cm^{-3}}$. 
\end{enumerate}

We perform 8 different runs, two of which do not undergo any particle refinement and are fixed at our low ($J_{\rm init} = 0.05$, similar to \citealt{Truelove1997}) and high ($J_{\rm init} = 0.01$) resolutions. These correspond to a mass resolution of $\Delta m = 3.77 \times 10^{-5} \, \Msun$ and $\Delta m = 2.95 \times 10^{-7} \, \Msun$ respectively. The remaining 6 runs undergo particle refinement from low resolution to high resolution. Refinement is done by keeping $J$ below a constant value, similar to many grid codes. Thus, whenever the Jeans number for a given cell/particle rises above this constant value, the particle is split to ensure the Jeans number criterion is satisfied. 

\begin{figure}
    \centering
    \includegraphics[width=\textwidth]{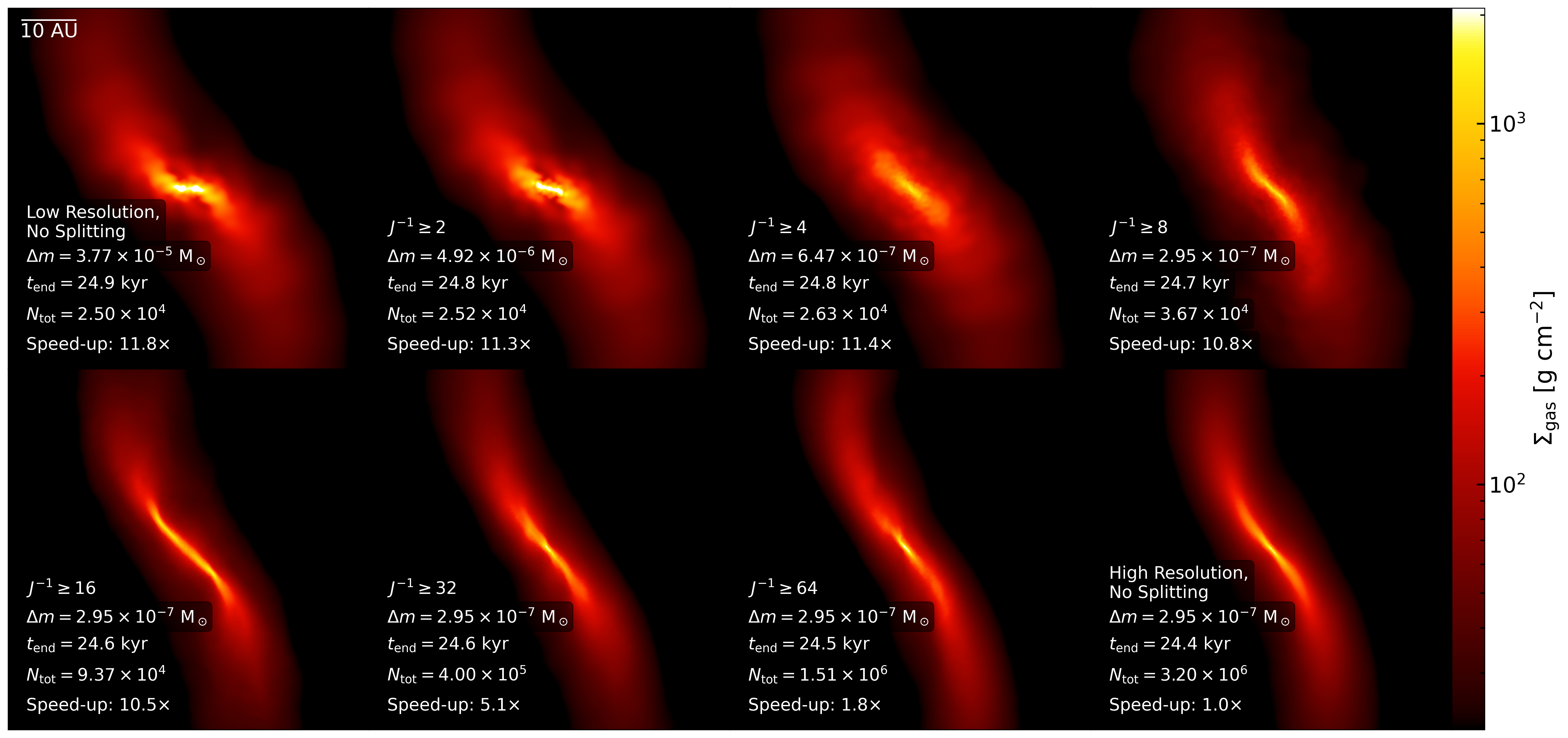}
    \caption{Tests for artificial fragmentation occurring while splitting particles. The top left and bottom right panels show the surface density at the end of the simulation ($\rho_{\rm max}=10^{-9.5} \, {\rm g \, cm^{-3}}$) in runs where we have uniformly (no particle splitting) low and high resolution respectively. The remaining panels show the final surface density in runs where we split particles to go from low to high resolution, while maintaining $J^{-1}\geq 2, 4, 8, 16, 32, 64$ respectively, where $J$ is the refinement Jeans number (\autoref{eq:refinement-jeans-number}). Symbols: $\Delta m$ indicates the minimum resolution achieved, $t_{\rm end}$ is the time taken to reach the end of the simulation and $N_{\rm tot}$ refers to the total number of particles present at the end of the simulation. The speed-up is calculated as the ratio of wall-clock times taken to reach the end for a run to that of the uniformly high resolution run. The formation of a single filament during collapse indicates that artificial fragmentation does not occur when we increase resolution by splitting particles.}
    \label{fig:jeans}
\end{figure}

In \autoref{fig:jeans}, we show the different runs at the end of the simulation. The top-left and bottom-right panels do not have any particle splitting and are fixed at our low and high resolution respectively. For the remaining \ps runs, we set $J^{-1} \geq 2,4,8,16,32,64$, to satisfy a minimum number of cells per Jeans length. 
The maximum Jeans number reached in the 8 different runs (in order of increasing number of total resolution elements) is $J_{\rm max} = 0.85, 0.43, 0.22, 0.17, 0.17, 0.17, 0.17,$ and $0.17$. \cite{Truelove1997} showed that if the Jeans condition is not satisfied, two centers of fragmentation form instead of one, in two different filamentary structures. We form only one filamentary structure in our runs, and therefore conclude that we do not see signatures of artificial fragmentation in our runs. However, we also note the presence of two over-densities in our low resolution run (without particle splitting), at a spacing of $\sim 2 \, {\rm AU}$. This is not surprising since the minimum cell size for this run is $0.41 \, {\rm AU}$ and $J_{\rm max} = 0.85$. Eventually, beyond a limit, the required gravitational resolution falls below the hydrodynamical resolution even in particle-like methods.

\bibliography{Refs}{}
\bibliographystyle{aasjournal}



\end{document}